%% file: main.tex
\documentclass[sigconf]{acmart}
\setcopyright{none} 
\acmConference[Anonymous Submission to WWW 2018]{The Web Conference}{Due Oct 31 2017}{Lyon, France}
\acmYear{2018}

\usepackage{booktabs}

\usepackage{epsfig,amsmath,multirow,makecell,caption,soul,csquotes,color,wrapfig,subcaption,mathtools,bm,upgreek,spverbatim}

\newenvironment{proof}{\paragraph{Proof:}}{\hfill$\square$}

\usepackage[boxed, ruled, vlined, linesnumbered]{algorithm2e}

\makeatletter
\providecommand{\leadsfrom}{%
  \mathrel{\mathpalette\reflect@squig\relax}%
}
\newcommand{\reflect@squig}[2]{%
  \reflectbox{$\m@th#1\leadsto$}%
}
\makeatother

\usepackage{graphicx}

\newenvironment{changemargin} [2]{\begin{list}{}{
          \setlength{\topsep}{0pt}\setlength{\leftmargin}{0pt}
          \setlength{\rightmargin}{0pt}
          \setlength{\listparindent}{\parindent}
          \setlength{\itemindent}{\parindent}
          \setlength{\parsep}{0pt plus 1pt}
          \addtolength{\leftmargin}{#1}\addtolength{\rightmargin}{#2}
          }\item }{\end{list}}

\newenvironment{myitemize} 
   {
     \begin{changemargin}{-3pt}{-0cm}
     \vspace{-10pt}
     \hspace{-5pt}
     \begin{itemize}
     \setlength{\itemsep}{3pt}
   }
   {
     \end{itemize}
     \vspace{-0pt}
     \end{changemargin}
   }

   {
     \begin{changemargin}{-8pt}{-0cm}
     \vspace{-13pt}
     \hspace{5pt}
     \begin{description}
     \setlength{\itemsep}{-1pt}
   }
   {
     \end{description}
     \end{changemargin}
   }

   {
     \begin{changemargin}{-8pt}{-0cm}
     \vspace{-13pt}
     \hspace{5pt}
     \begin{enumerate}
     \setlength{\itemsep}{1pt}
   }
   {
     \end{enumerate}
     \end{changemargin}
   }

\newcommand{\myref}[1]{\S\,\ref{#1}}

\newcommand{\dl}{{{DL}}\xspace}
\newcommand{\dd}{{{DD}}\xspace}

\newcommand{\ar}{{{AR}}\xspace}
\newcommand{\arp}{{{ARP}}\xspace}
\newcommand{\ars}{{{ARs}}\xspace}
\newcommand{\pv}{{{PV}}\xspace}
\newcommand{\pvs}{{{PVs}}\xspace}

\newcommand{\mar}{{{ARA}}\xspace}
\newcommand{\ara}{{{ARA}}\xspace}

\newcommand{\dnn}{{{DNN}}\xspace}
\newcommand{\dnns}{{{DNNs}}\xspace}

\newcommand{\cnn}{{{\sc Cnn}}\xspace}
\newcommand{\mxn}{{{\sc Mxn}}\xspace}
\newcommand{\nin}{{{\sc Nin}}\xspace}

\newcommand{\svhn}{{{\sc Svhn}}\xspace}
\newcommand{\cifar}{{{\sc Cifar10}}\xspace}
\newcommand{\mnist}{{{\sc Mnist}}\xspace}
\newcommand{\glm}{{{\sc Glm}}\xspace}
\newcommand{\mlp}{{{\sc Mlp}}\xspace}

\newcommand{\ttg}{{ {\sc G-}}\xspace}
\newcommand{\tth}{{ {\sc H-}}\xspace}
\newcommand{\ttp}{{ {\sc P-}}\xspace}

\newcommand{\ttga}{{ {\sc G-A}}ttack\xspace}
\newcommand{\ttha}{{ {\sc H-A}}ttack\xspace}
\newcommand{\ttpa}{{ {\sc P-A}}ttack\xspace}
\newcommand{\ttca}{{ {\sc C-A}}ttack\xspace}
\newcommand{\ttgt}{{ {\sc G-}}trained\xspace}

\newcommand{\ttct}{{ {\sc C-}}trained\xspace}

\makeatletter
\newcommand{\vo}{\vv{o}\@ifnextchar{)}{\,}{}}
\makeatother

\makeatletter
\newcommand{\vr}{\vv{r}\@ifnextchar{)}{\,}{}}
\makeatother

\newcommand{\mbs}{\sigma}
\newcommand{\mbsp}{\sigma_\epsilon}
\newcommand{\mbw}{w}
\newcommand{\mbr}{r}
\newcommand{\mbg}{\pi}
\newcommand{\mbyp}{o_\epsilon}
\newcommand{\mby}{o}
\newcommand{\mbx}{x}
\newcommand{\mbxp}{x_{\epsilon}}
\newcommand{\mbxy}{x_*}

\newcommand{\mbxpp}{x_{\epsilon^2}}

\newcommand{\mbz}{z}

\newcommand{\mbJ}{J}

\newcommand{\system}{{EagleEye}\xspace}

\newtheorem*{theorem*}{Theorem}

\begin{document}

\title{EagleEye: Attack-Agnostic Defense against Adversarial Inputs\\
(Technical Report)}

\author{Yujie Ji \qquad Xinyang Zhang \qquad Ting Wang}
\affiliation{%
  \institution{Lehigh University}
  \city{Bethlehem, PA}
}
\email{{yuj216, xizc15}@lehigh.edu, inbox.ting@gmail.com }

\input{abstract.tex}



\keywords{Machine learning system; Adversarial input; Attack-agnostic defense}

\maketitle


\input{introduction.tex}

\input{background.tex}

\input{measurement.tex}
\input{model.tex}

\input{analysis.tex}
\input{evaluation.tex}
\input{discussion.tex}
\input{literature.tex}
\input{conclusion.tex}

\newpage

\bibliographystyle{acm}
\bibliography{main}

\input{appendix.tex}

\end{document}

%% file: abstract.tex
\begin{abstract}

Deep neural networks (\dnns) are inherently vulnerable to adversarial inputs: such maliciously crafted samples trigger \dnns to misbehave, leading to detrimental consequences for \dnn-powered systems. The fundamental challenges of mitigating adversarial inputs stem from their adaptive and variable nature. Existing solutions attempt to improve \dnn resilience against specific attacks; yet, such static defenses can often be circumvented by adaptively engineered inputs or by new attack variants.

Here, we present \system, an attack-agnostic adversarial tampering analysis engine for \dnn-powered systems. Our design exploits the {\em minimality principle} underlying many attacks: to maximize the attack's evasiveness, the adversary often seeks the minimum possible distortion to convert genuine inputs to adversarial ones. We show that this practice entails the distinct distributional properties of adversarial inputs in the input space. By leveraging such properties in a principled manner, \system effectively discriminates adversarial inputs and even uncovers their correct classification outputs. Through extensive empirical evaluation using a range of benchmark datasets and \dnn models, we validate \system's efficacy. We further investigate the adversary's possible countermeasures, which implies a difficult dilemma for her: to evade \system's detection, excessive distortion is necessary, thereby significantly reducing the attack's evasiveness regarding other detection mechanisms.
\end{abstract}

%% file: introduction.tex
\section{introduction}
\label{sec:introduction}

Recent years have witnessed the abrupt advances in deep learning (\dl) techniques~\cite{LeCun:2015:nature}, which lead to breakthroughs in a number of long-standing artificial intelligence tasks (e.g., image classification, speech recognition, and even playing Go~\cite{go}). Internet giants, such as Google, Facebook and Amazon, all have heavily invested in offering \dl-powered services and products.

\begin{figure}
    \centering
    \epsfig{width=80mm, file=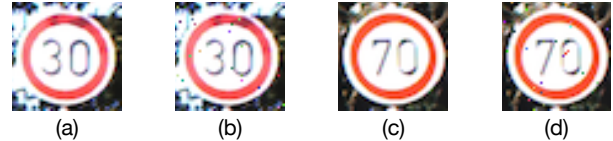}
    \vspace{-5pt}
    \caption{(a) (c) genuine inputs - both are correctly recognized; (b) (d) adversarial inputs - (b) is misclassified as ``70 mph'' and (d) is misclassified as ``30 mph''.\label{fig:adv}}
\end{figure}

However, designed to model highly nonlinear, nonconvex functions, deep neural networks (\dnns) are inherently vulnerable to adversarial inputs, which are malicious samples crafted by adversaries to trigger \dnns to misbehave~\cite{Szegedy:2013:arxiv}. Figure~\ref{fig:adv} shows an example: both original images are correctly recognized by a \dnn; with a few pixels altered, the resulting adversarial images are misclassified by the same \dnn, though the difference is barely discernible for human eyes. With the increasing use of \dl-powered systems in security-critical domains, adversaries have strong incentive to manipulate such systems via forcing misclassification of inputs: illegal content can bypass content filters that employ \dl to discriminate inappropriate web
content~\cite{Grosse:arxiv:2016}; biometric authentications that apply \dl to validate human faces can be manipulated to allow improper access~\cite{Sharif:2016:ccs}; in the near future, driverless vehicles that use \dl to detect traffic signs may be misled to crashing.

The fundamental challenges of defending against adversarial input attacks stem from their adaptive and variable nature: they are created tailored to target \dnns, while crafting strategies vary greatly with concrete attacks. Existing solutions attempt to improve \dnn resilience against specific attacks~\cite{Gu:2014:arxiv,Goodfellow:2014:arxiv,Huang:2015:arxiv,Shaham:2015:arxiv,Papernot:2016:sp}; yet, such static defenses, once deployed, can often be circumvented by adaptively engineered inputs or by new attack variants. For instance, the training data augmentation mechanism~\cite{Goodfellow:2014:arxiv,Nokland:2015:arXiv} suggests to train \dnns on adversarial inputs; as detailed in~\myref{sec:measure}, the resulting models often overfit to known attacks, thus being even more vulnerable to unseen variants. Further, most existing solutions require significant modifications to either \dnn architectures or training procedures, which often negatively impact the classification accuracy of \dnn models. Indeed, recent theoretical exploration~\cite{Fawzi:2015:arxiv} has confirmed the inherent trade-off between \dnn robustness and expressivity, which significantly impedes the adoption of existing defense solutions in accuracy-sensitive domains.

In this paper, we take a completely new route: instead of striving to improve \dnn robustness against specific attacks, we aim at defense mechanisms that make minimal assumptions regarding the attacks and adapt readily to their variable nature. To this end, we design, implement and evaluate \system, an attack-agnostic adversarial tampering analysis engine for \dl-powered systems.

At a high level, \system leverages the {\em minimality principle} underlying many attacks: intuitively, to maximize the attack's evasiveness, the adversary often seeks the minimum possible distortion to convert a genuine input to an adversarial one. We show both empirically and analytically that this practice entails the distinct properties shared by adversarial inputs: compared with their genuine counterparts, adversarial inputs tend to distribute ``closer'' to the classification boundaries induced by \dnns in the input manifold space.
%
By exploiting such properties in a principled manner, \system effectively discriminates adversarial inputs and even uncovers their correct classification outputs. We also investigate the adversary's possible countermeasures by abandoning the minimality principle, which however implies a difficult dilemma for her: to evade \system's detection, excessive distortion is necessary, thereby significantly reducing the attack's evasiveness with respect to other detection mechanisms (e.g., human vision).

Note that we are not arguing to replace existing defense solutions with \system. Rather, their distinct designs entail their complementary nature. \system exerts minimal interference to existing components of \dl-powered systems and is thus compatible with existing defenses. Moreover, the synergistic integration of \system with other mechanisms (e.g., defensive distillation~\cite{Papernot:2016:sp}) delivers even stronger defenses for \dnns.

Our contributions can be summarized as follows.
\begin{itemize}
    \item We expose the limitations of existing defenses against adversarial input attacks, which motivates the design of \system. To our best knowledge, our empirical evaluation (\myref{sec:measure}) is the most comprehensive study to date on varied attack and defense models.

    \item We identify the minimality principle underlying most attacks, which entails the distinct properties shared by adversarial inputs. We design and implement \system, which effectively exploits such properties in a principled manner (\myref{sec:model}).

    \item We analytically and empirically validate \system's efficacy (\myref{sec:analysis}  and \myref{sec:evaluation}), which achieves promising accuracy in discriminating adversarial inputs and even uncovering their correct classification outputs.

    \item We investigate the adversary's possible countermeasures and their implications. We also empirically explore the synergistic integration of \system with existing defense mechanisms (\myref{sec:evaluation} and \myref{sec:discussion}).

\end{itemize}

All the source code of this paper will be released on {\sf GitHub} after the double-blind review is complete.

%% file: background.tex
\begin{figure}
\centering
\epsfig{file=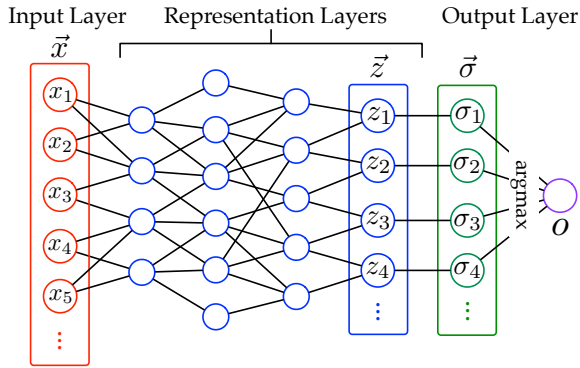, width=80mm}
\caption{Illustration of \dnn models. \label{fig:network}}
\end{figure}

\section{Attacks and Defenses}
\label{sec:background}

In this section, we introduce a set of fundamental concepts and assumptions, and survey representative attack and defense models in literature.

\subsection{Deep Learning}

\dl represents a class of machine learning algorithms designed to learn high-level abstraction of complex data using multiple processing layers and nonlinear transformations. Figure~\ref{fig:network} shows a typical \dnn architecture.

In this paper, we primarily focus on image classification tasks, while our discussion generalizes to other settings (see \myref{sec:discussion}). In these tasks, the \dnn encodes a mapping $f: \mathcal{X} \rightarrow \mathcal{O}$,  which assigns a given image $\mbx$ (represented as a vector) in the input space ${\mathcal X}$ to one of a set of classes ${\mathcal O}$. For example, with the \mnist dataset as inputs, $f$ classifies each image as one of ten digits `0'-`9'. As shown in Figure~\ref{fig:network}, the last layer of $f$ often employs a softmax function. Specifically, let
\begin{displaymath}
 \vec{z}  \triangleq  \left[z_1, z_2, \ldots \right]\qquad \mbs  =  \frac{\exp(\mbz)}{\sum_i \exp(z_i)} \triangleq \left[\sigma_1, \sigma_2, \ldots \right]
\end{displaymath}
 respectively be the input and output of this layer. Then $\sigma_i$ is the probability that $\mbx$ belongs to the $i^{\rm th}$ class. The predicted class of $\mbx$ is given by $f(\mbx) = \arg\max_i \sigma_i$.

We consider \dnns obtained via supervised learning. Specifically, to train a \dnn $f$, the algorithm takes a training set, of which each instance $(\mbx, o) \in {\mathcal X} \times {\mathcal O}$ constitutes an input and its ground-truth class, and determines the parameter setting of $f$ via minimizing a loss function $\ell(f(\mbx), o)$ (e.g., the cross entropy of ground-truth classes and $f$'s outputs).

\subsection{Attack Methodologies}
\label{sec:attack}

Targeting a \dnn $f$ deployed in use, the adversary attempts to trigger $f$ to misbehave by feeding it with carefully crafted inputs. Given a genuine input $\mbx$ correctly classified by $f$, the adversary generates an adversarial one $\mbxp$ by perturbing $\mbx$ with an insignificant amplitude (e.g., a few pixels). The difference of $\mbx$ and $\mbxp$ ($\mbr = \mbxp - \mbx$) is referred to as the {\em perturbation vector} (\pv).

We differentiate two attack scenarios. In an untargeted attack, the adversary is interested in simply forcing $f$ to misclassify, i.e., $f(\mbxp)\neq f(\mbx)$. In a targeted attack, she further desires for a particular target output $\mbyp$, i.e., $f(\mbxp) = \mbyp$. In the following we focus our discussion on targeted attacks, while the extension to untargeted attacks is straightforward.
%

A variety of attack models have been proposed in literature~\cite{Goodfellow:2014:arxiv,Huang:2015:arxiv,Papernot:2016:eurosp,Carlini:2016:arXiv}. Despite their variations in concrete crafting strategies, they all roughly follow a two-step procedure: (i) {\em saliency estimation:} the adversary assesses the impact of changing each input component on the classification output; (ii) {\em perturbation selection:} the adversary uses the input saliency information to select and perturb a subset of input components. Based on the concrete implementation of the two steps,  we classify existing attacks in two major categories.

\subsubsection{Linear Crafting Attacks}
The class of linear attacks estimate the impact of distorting different input components on $f$'s output via linear approximations and find the \pv $\mbr$ that maximizes the probability of the target output $\mbyp$. Next we detail two representative attack models.

\paragraph{Goodfellow's Attack.\;} Goodfellow {\em et al.}~\cite{Goodfellow:2014:arxiv} proposed the first linear attack model, which computes the gradient of the loss function $\ell$ with respect to the input $\mbx$ and determines $\mbr$ as a gradient sign step in the direction that increases the probability of the target output $\mbyp$.

Specifically, let $\textrm{sign} (\triangledown_{\mbx} \ell (f(\mbx), \mbyp) )$ be the gradient sign of $\ell$ with respect $\mbx$ for given $\mbyp$. Then the \pv is defined as:
$\mbr = - \delta  \cdot \textrm{sign} (\triangledown_{\mbx} \ell (f(\mbx), \mbyp) )$,
where $\delta$ is a parameter controlling the distortion amplitude (i.e., $l_\infty$-norm of $\mbr$). Often, the adversary seeks the minimum $\delta$ to achieve misclassification:
$\min_\delta f( \mbx + \mbr) = \mbyp$.

\paragraph*{Huang's Attack.\;} Huang {\em et al.}~\cite{Huang:2015:arxiv} introduced another linear attack model, which is constructed upon a linear approximation of the output of the softmax layer, i.e., the last layer of a \dnn model (see  Figure~\ref{fig:network}).

Let $\mbsp$ be the softmax output of $\mbxp = \mbx + \mbr$. This attack approximates $\mbsp$ using a linear form:
$\mbsp \approx \mbs + \mbJ \mbr$,
 where $\mbJ = \frac{{\rm d} \mbs}{{\rm d} \mbx}$ is the Jacobian matrix.
Assume the original output $\mby$ and target output $\mbyp$ respectively correspond to the $j^{\rm th}$ and $j_\epsilon^{\rm th}$ row of $J$, denoted by $J_j$ and $J_{j_\epsilon}$. Let $\Delta_J \triangleq J_{j_\epsilon}  - J_j$. To trigger $f(\mbxp) = \mbyp$, the adversary seeks $\mbr$ that maximizes the difference of the $j_\epsilon^{\rm th}$ and $j^{\rm th}$ component of $\mbsp$, i.e.,
$\max_{\mbr}  \Delta_J  \cdot \mbr$.

Similar to~\cite{Goodfellow:2014:arxiv}, this attack determines $\mbr$ as a step in the sign direction of $\Delta_J$, i.e.,
$ \mbr = \delta \cdot  \textrm{sign}(\Delta_J)$,
where $\delta$ controls the distortion amplitude\footnote{In~\cite{Huang:2015:arxiv} Huang {\em et al.} also give the definition of optimal $\mbr$ when the distortion amplitude is measured by $l_1$- or $l_2$-norm of $\mbr$.}.

\subsubsection{Nonlinear Crafting Attacks}

In linear attacks, the \pvs are found in a single attempt. In comparison, nonlinear attacks construct the \pvs iteratively. At each round, the adversary estimates the impact of each input component on the classification output, then selects several components to perturb, and checks whether the updated input causes the target misclassification.
Next we detail two representative nonlinear attacks.

\paragraph*{Papernot's Attack.\;} Papernot {\em et al.}~\cite{Papernot:2016:eurosp} proposed a saliency map to guide the crafting process. Intuitively, this map describes the impact of each input component on the output. Given the current input $\mbx$ (with previously selected components perturbed) and target output $\mbyp$ (corresponding to the $j_\epsilon^{\rm th}$ component of $\mbs$), the $i^{\rm th}$ component is associated with a pair of measures:
\begin{displaymath}
\alpha_i  =  \frac{\partial \sigma_{j_\epsilon}}{\partial x_i} \qquad
\beta_i  =   \sum_{ j \neq j_\epsilon }\frac{\partial \sigma_j}{\partial x_i}
\end{displaymath}
where $\alpha_i$ is its impact on the probability of $\mbyp$, while $\beta_i$ is its impact on all the other classes.

The attack consists of multiple iterations of a greedy procedure.
At each round, two components with the largest value of $(-\alpha \cdot \beta)$ are selected and flipped (to either `$1$' or `$-1$'), and the saliency map is updated accordingly. This process continues until the resulting input is misclassified as $\mbyp$. The distortion amplitude is defined by the number of distorted components, i.e., $l_1$-norm of $\mbr$.


\paragraph*{Carlini's Attack.\;} In response to the defensive distillation method~\cite{Papernot:2016:sp}, Carlini and Wagner introduced a nonlinear attack~\cite{Carlini:2016:arXiv}, which differs from~\cite{Papernot:2016:eurosp} in two  aspects:

\begin{itemize}
\item To compensate for the gradient vanishing due to defensive distillation, the input to the softmax layer is artificially amplified by $\tau$ times, where $\tau$ is the ``temperature'' used by defensive distillation. The output of the softmax layer is thus:
$\mbs = \exp(\frac{\mbz}{\tau})/\sum_j \exp(\frac{z_j}{\tau})$.

\item The saliency values of input components are defined as $|\alpha - \beta|$ rather than $(-\alpha \cdot  \beta)$. This modification reduces the complexity of perturbation selection from $O(n^2)$ to $O(n)$, where $n$ is the number of input components. At each iteration, a pair of input components with the largest saliency values are selected and flipped.
\end{itemize}

\subsubsection{Linear vs. Nonlinear Attacks}

Linear attacks require computing gradient or Jacobian only once, while nonlinear attacks often involve multiple rounds of gradient or Jacobian computation. Given their efficiency advantage, linear attacks can be exploited to craft a large number of adversarial inputs.

Meanwhile, existing linear attacks often measure the distortion amplitude by $l_\infty$-norm of the \pv, while existing nonlinear attacks attempt to minimize $l_1$-norm or $l_2$-norm of the \pv.

The empirical comparison of the characteristics of different attacks is detailed in~\myref{sec:measure}.

\subsection{Defense Methodologies}
\label{sec:defense}

A \dnn's resilience against adversarial inputs is inherently related to its stability~\cite{Bastani:2016:arXiv}. Intuitively, a \dnn $f$ is stable, if for any ``proximate'' inputs $\mbx$ and $\mbxp$ (i.e., $||\mbxp - \mbx||$ is small), $f(\mbx)$ and $f(\mbxp)$ are similar. Motivated by this rationale, a plethora of solutions~\cite{Gu:2014:arxiv,Goodfellow:2014:arxiv,Huang:2015:arxiv,Shaham:2015:arxiv,Papernot:2016:sp} have been proposed to improve \dnn stability, which can be roughly classified in three major categories.

\paragraph*{Data Augmentation.\;}
This class of methods improve \dnn stability by proactively generating a set of adversarial inputs and incorporating them in the training process. Formally, given a \dnn $f$ and a known attack $g$, via applying $g$ over $f$, one generates an adversarial input $\mbxp$ for each genuine instance $(\mbx, \mby)$. A new \dnn $f'$ is trained using an augmented objective function:
\begin{equation*}
\min_f \sum_{(\mbx, \mbxp, \mby)}  (\alpha \cdot \ell(f(\mbx), \mby) + (1-\alpha) \cdot \ell(f(\mbxp), \mby))
\end{equation*}
where the parameter $\alpha$ $(0 \leq \alpha \leq 1)$ balances the relative weight of genuine and adversarial inputs.
For instance, Goodfellow {\em et al.}~\cite{Goodfellow:2014:arxiv} suggested equal importance of genuine and adversarial inputs ($\lambda = 0.5$),

Nevertheless, these methods are inherently heuristic, without theoretical guarantee on the robustness or accuracy of the trained \dnn models.

\paragraph*{Robust Optimization.\;}

Another line of work proposed to improve \dnn stability via directly altering its objective function. To be specific, one prepares a \dnn $f$ for the worst possible inputs by training it with an minimax objective function:
\begin{equation}
    \label{eq:rof}
  \min_f  \max_{ ||\mbr || \leq \delta} \ell(f(\mbx + \mbr), o)
\end{equation}

The training algorithm first searches for the ``worst'' \pv (constrained by $\delta$) that maximizes the loss function $\ell$ under the current setting of $f$; it then optimizes $f$ with respect to this \pv. This objective function essentially captures the misclassification error under adversarial perturbations.

Due to the complexity of \dnn models, it is intractable to search for exact worst adversarial inputs. Certain simplifications are often made. Szegedy {\em et al.}~\cite{Szegedy:2013:arxiv} and Gu and Rigazio~\cite{Gu:2014:arxiv} proposed to search for $\mbr$ along the gradient direction of loss function, while Shaham {\em et al.}~\cite{Shaham:2015:arxiv} and Miyato {\em et al.}~\cite{Miyato:2015:arXiv} reformulated this framework for Kullback-Leibler divergence like loss functions.

\paragraph*{Model Transfer.\;} In this method, one transfers the knowledge of a teacher \dnn $f$ to a student \dnn $f'$ such that the model stability is improved.

For instance, Papernot {\em et al.}~\cite{Papernot:2016:sp} proposed to employ distillation~\cite{Ba:2014:nips,Hinton:2015:arxiv}, a technique previously used to transfer the knowledge of an ensemble model into a single model, to improve \dnn stability. Specifically,
\begin{itemize}
\item The teacher \dnn $f$ is trained on genuine inputs; in particular, the input to the softmax layer (see Figure~\ref{fig:network}) is modified as $z/\tau$  for given ``temperature'' $\tau (\tau > 1)$.
\item One evaluates $f$ on the training set and produces a new training set $\{(\mbx, \mbs)\}$, where $\mbs$ encodes the predicted probability distribution (``soft label'') of $\mbx$.
\item By training the student \dnn $f'$ on the new training set under temperature $\tau$, $f'$ is expected to generalize better to adversarial inputs than $f$.
\end{itemize}

Additionally, there is recent work attempting to design new \dnn architectures~\cite{Gu:2014:arxiv} or learning procedures~\cite{Chalupka:2014:arXiv} to improve \dnn stability. Yet, the resulting models fail to achieve satisfying accuracy on genuine inputs. Due to space limitations, we focus our discussion on the above three classes of defense mechanisms.

%% file: measurement.tex
\begin{table*}{\small
    \centering
\begin{tabular}{|c|c|c|c|c|c|c|c|c|}
  \hline
  \multirow{3}{*}{\bf Data}    & \multirow{3}{*}{\bf Original Model} & \multicolumn{7}{c|}{\bf Defense-Enhanced Models}\\
   \cline{3-9}
& &   \multicolumn{4}{c|}{\bf Data Augmentation ($\alpha = 0.5$)} & \multicolumn{2}{c|}{\bf Robust Optimization} & {\bf Model Transfer} \\ \cline{3-8}
    &    & {\bf G-trained} & {\bf H-trained} & {\bf P-trained} & {\bf C-trained} & {\bf $l_1$-norm} & {\bf $l_\infty$-norm} &  {\bf ($\tau = 40$)}\\
    \hline
    \hline
\mnist  &   99.5\% & 98.8\% & 99.0\% & 99.0\% & 98.8\% & 98.1\%& 98.5\%  & 98.9\% \\
\cifar & 85.2\% & 64.4\% & 57.6\% & 75.9\% & 76.7\% & 72.1\%& 71.3\% & 80.5\% \\
\svhn & 95.2\% & 91.3\% & 85.0\% & 91.2\% & 92.4\% & 90.2\%& 81.5\% & 86.0\% \\
    \hline
\end{tabular}
\caption{Classification accuracy of original and defense-enhanced \dnn models with respect to benchmark datasets.  \label{tab:accuracy}}}
\end{table*}


\begin{table*}
    \centering
    {\small
\begin{tabular}{|c|c|c|c|c|c|c|c|c|c|}
  \hline
  \multirow{3}{*}{\bf Data}  & \multirow{3}{*}{\bf Attack}  & \multirow{3}{*}{\bf Original Model} & \multicolumn{7}{c|}{\bf Defense-Enhanced Models}\\
   \cline{4-10}
& & &  \multicolumn{4}{c|}{\bf Data Augmentation ($\alpha = 0.5$)} &  \multicolumn{2}{c|}{\bf Robust Optimization} & {\bf Model Transfer} \\ \cline{4-9}
    &  &  & {\bf G-trained} & {\bf H-trained} & {\bf P-trained} & {\bf C-trained} &  {\bf  $l_1$-norm} &  {\bf  $l_\infty$-norm} &  {\bf ($\tau = 40$)}\\
    \hline
    \hline
   \multirow{4}{*}{\rotatebox{90}{\mnist}}
   & {\bf G-} &15.3\%  & 5.9\%  & 82.4\%  & 40.0\%  & 70.6\%  & 21.2\% & 0.0\%  & 1.18\%  \\
   & {\bf H-} &22.4\%  & 7.1\%  & 87.1\%  & 84.7\%  & 91.8\%  & 22.4\% & 0.0\% & 1.18\%  \\
   & {\bf P-} &{\bf 100.0\%}  & {\bf 100.0\%}  & {\bf 100.0\% } & {\bf 100.0\% } & {\bf 100.0\% } & {\bf 100.0\%}  & {\bf 100.0\%}  & 1.0\%  \\
   & {\bf C-} & 100.0\%  & 100.0\%  & 100.0\%  & 100.0\%  & 100.0\%  &  100.0\% & 100.0\%  & {\bf 100.0\% } \\
    \hline
    \hline
    \multirow{4}{*}{\rotatebox{90}{\cifar}}
    & {\bf G-} &96.5\%  & {\bf 100.0\%}  & {\bf 100.0\%}  & 93.3\%  & {\bf 100.0\%}  & {\bf 98.1\%} & 79.7\%  & 39.8\%  \\
    & {\bf H-} &91.9\%  & 100.0\%  & 100.0\%  & {\bf 100.0\%}  & 93.8\%  & 98.1\% & 71.2\%  & 38.6\%  \\
    & {\bf P-} & {\bf 100.0\%}  &  100.0\%  &  100.0\%  &  100.0\%  &  100.0\%  & 94.3\%  & {\bf 100.0\% } & 21.7\%  \\
    & {\bf C-} & 100.0\%  & 100.0\%  & 100.0\% & 100.0\%  & 100.0\%  & 64.2\% & 98.31\%   & {\bf 100.0\% } \\
     \hline
     \hline
     \multirow{4}{*}{\rotatebox{90}{\svhn}}
     & {\bf G-} &99.5\%  & 36.9\%  & 4.39\%  & 99.5\% & 99.5\%  & 100.0\% & 0.0\%  & 7.5\%  \\
     & {\bf H-} &94.6\%  & 35.4\%  & 5.37\%  & 93.4\%  & 94.9\%  & 96.7\%  & 0.0\%  & 7.5\%  \\
     & {\bf P-} & {\bf 100.0\% } & {\bf 100.0\% } & {\bf 100.0\% } & {\bf 100.0\% } & {\bf 100.0\% } & {\bf 100.0\%} & {\bf 99.7\% } & 3.0\% \\
     & {\bf C-} &100.0\%  & 98.0\%  & 98.1\%  & 100.0\% & 99.5\%  & 95.6\%   & 90.8\% & {\bf 100.0\% } \\
      \hline

\end{tabular}}
\caption{Resilience of original and defense-enhanced \dnn models against adversarial input attacks.  \label{tab:robustness}}
\end{table*}

\section{Empirical Study}
\label{sec:measure}

Next we empirically evaluate the effectiveness of existing defense solutions against varied attacks. To our best knowledge, this evaluation represents the most comprehensive study to date on a range of attack and defense models, and is thus interesting in its own right.

In a nutshell, we show that it is fundamentally challenging to defend against adversarial inputs, which are tailored to target \dnns and crafted with varying strategies.
Unfortunately, existing defenses are inherently static. Although they improve \dnn resilience against specific attacks, the resulting models, once trained and deployed, are unable to adapt to {\em a priori} unknown attacks. The adversary can thus circumvent such defenses by creating inputs exploiting new vulnerability of target \dnns.



\subsection{Setting of Study}

\paragraph*{Datasets and DNN Models.\;}
To show the prevalence of attack vulnerabilities across different tasks, in our study, we use three benchmark datasets, \mnist~\cite{mnist},  \cifar~\cite{cifar}, and \svhn~\cite{svhn}, which have been widely used to evaluate image classification algorithms. The details of datasets can be found in Appendix A.

We also consider three distinct \dnn architectures and apply each to one of the datasets above. To be specific, we apply the convolutional neural network (\cnn)~\cite{LeCun:1998:cnn}, maxout network (\mxn)~\cite{lin:2014:iclr}, and network-in-network (\nin)~\cite{lin:2014:iclr} models to classifying the \mnist, \cifar, and \svhn datasets, respectively. The implementation details of these \dnn models are referred to Appendix B.


%
%
%
%
%


\paragraph*{Attacks and Defenses.\;} We implement all the attack models in~\myref{sec:attack}, which we refer to as \ttg, \tth, \ttp, and \ttca for brevity. In particular, following~\cite{Goodfellow:2014:arxiv,Huang:2015:arxiv}, we set the limit of distortion amplitude for \ttg and \ttha as 0.25 (i.e., $\delta \leq 0.25$ in~\myref{sec:attack}); for \ttp and \ttca, as in~\cite{Papernot:2016:sp}, we fix this limit to be 112, i.e., the adversary is allowed to perturb no more than 112 pixels.

We implement one representative solution from each defense category in~\myref{sec:defense}. In particular, as data augmentation is attack-specific, we refer to the resulting models as \ttg, \tth, \ttp, and {\ttct} \dnn; as robust optimization is norm-specific, we train robust \dnns under both $l_\infty$- and $l_1$-norm criteria. The implementation details of defense methods are referred to Appendix C.

\subsection{``No Free Lunch''}

Table~\ref{tab:accuracy} summarizes the classification accuracy of the original \dnn models (\cnn, \mxn, \nin) trained over legitimate inputs and their defense-enhanced variants on the benchmark datasets (\mnist, \cifar, \svhn).


 Observe that the original models achieve accuracy (i.e., 99.5\%, 85.2\%, 95.2\%) close to the state of the art~\cite{classification:result}. In comparison, most of their defense-enhanced variants observe non-trivial accuracy drop. For example, the accuracy decreases by 4.7\% from the original \mxn model to its defensive distilled variant (model transfer), while this drop is as significant as 20.8\% in the case of the \ttgt variant (data augmentation).

We thus conclude that the improvement of attack resilience is not ``free lunch'', often at the expense of classification accuracy. This observation is consistent with the theoretical investigation on the trade-off between \dnn expressivity and robustness~\cite{Fawzi:2015:arxiv}.

\subsection{``No Silver Bullet''}

Next we evaluate different \dnns' attack resilience. Under the limit of distortion amplitude, we measure the percentage of legitimate inputs in each testing set which can be converted to adversarial inputs by varied attacks. Table~\ref{tab:robustness} summarizes the results. The most successful attack under each setting is highlighted.

For the original models, most attacks, especially \ttp and \ttca, achieve near-perfect success rates, implying the prevalence of vulnerabilities across \dnn models.

The data augmentation method significantly improves \dnn resilience against linear attacks. The success rate of \ttga drops below  6\% when facing {\ttgt} \cnn.
However, it is much less effective for more complicated \dnns or against nonlinear attacks. Both \ttp and \ttca achieve near-perfect success rates against data augmented \mxn and \nin. This is because data augmentation is only capable of capturing simple, linear perturbations, while the space of \pvs for nonlinear attacks and complex \dnn models is much larger.

By considering worst-case inputs at every step of training, the robust optimization method leads to stronger resilience against linear attacks. For example, in the cases of \mnist and \svhn, the enhanced \dnn models completely block \ttg and \ttha.
 However, similar to data augmentation, robust optimization is ineffective against nonlinear attacks. This is partially explained by that the adversarial perturbations considered in training are essentially still linear (see Eq.(\ref{eq:rof2})).

Model transfer is the only defense effective against nonlinear attacks. The success rate of \ttpa drops to 1\% against defensive distilled \cnn, which is consistent with the results in~\cite{Papernot:2016:sp}. However, this effectiveness is not universal. 
\ttca, which is engineered to negate the gradient vanishing effects, is able to penetrate the protection of defensive distillation completely.

From the study above, we conclude that none of the existing defense solutions is a ``silver bullet''. While they improve \dnn resilience against specific attacks, the resulting models are unable to adapt to new attack variants. There is thus an imperative need for attack-agnostic defense mechanisms.

%% file: model.tex
\section{Defending DNN with EagleEye}
\label{sec:model}

\begin{figure}
  \epsfig{file=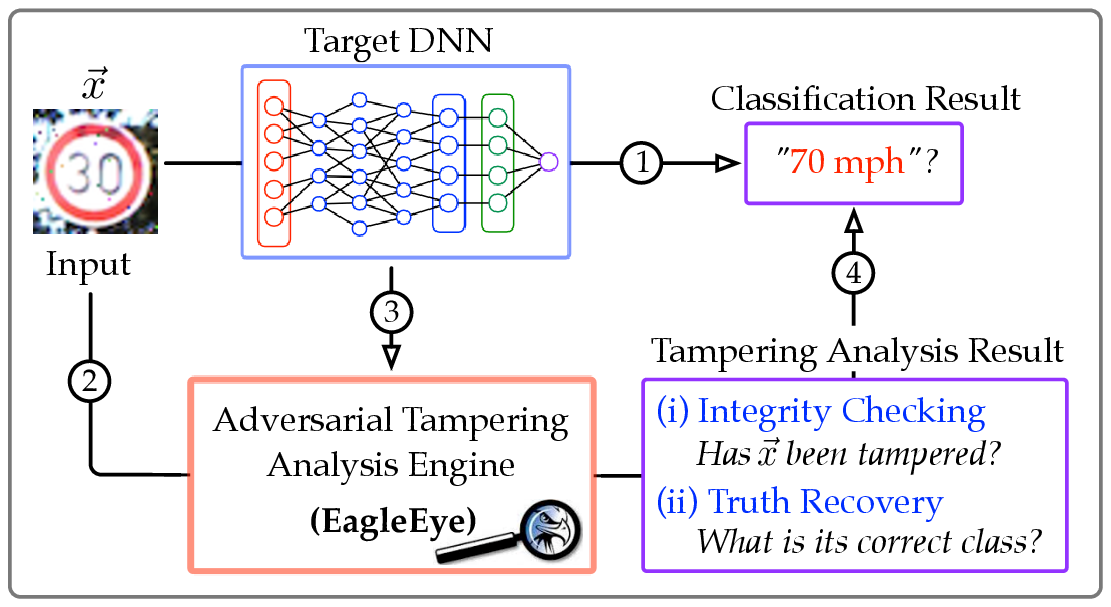, width=90mm}
  \caption{Use case of \system. \label{fig:arch}}
\end{figure}

Next we present \system, an attack-agnostic adversarial tampering analysis engine. Its design is motivated by a set of desiderata, which we believe are expected for practical and effective defense mechanisms.

\begin{itemize}

\item Attack-agnostic defense. It should be universally effective against known and {\em a priori} unseen attacks.

\item Intact \dnn models. It should require no modification to \dnn models, as such changes, especially to \dnn architectures or training procedures, often result in unpredictable system behaviors.



\item Light-weight execution. It should incur negligible performance overhead to the \dl-powered system.

\end{itemize}
%

\system satisfies all these desiderata. In its overview (\myref{sec:overview}), we show that \system, following a modular design, requires no modification to \dnn models or training methods; in its detailed description (\myref{sec:adefense}), we demonstrate that \system provides universal defense by making minimal assumptions regarding incoming attacks; in~\myref{sec:evaluation}, we further validate that \system meets the requirement of light-weight execution by evaluating its empirical performance within \dl-powered systems.

\subsection{Overview}
\label{sec:overview}

In contrast of existing solutions, \system takes a completely new route: it attempts to discriminate adversarial inputs; moreover, for suspicious cases, it is instrumented to infer their correct classification outputs. Therefore, \system provides much richer diagnosis information than existing defense solutions.



Specifically, as depicted in Figure~\ref{fig:arch}, \system is deployed as an auxiliary module within a \dl-powered system. It exerts minimal interference with the classification task (1). Rather, for the given input $\mbx$ (2) and \dnn (3), it offers on-demand adversarial tampering analysis: (i) it first runs {\em integrity checking} to assess the possibility that $\mbx$ has been maliciously tampered; (ii) if suspicious, it further performs {\em truth recovery} to infer $\mbx$'s correct classification. The analysis result is combined with the \dnn's classification to form a comprehensive report for the operator's decision-making (4). Clearly the design of \system satisfies the desiderata of intact \dnn models. Next we focus on realizing attack-agnostic defense.


\subsection{Minimality Principle}
\label{sec:adefense}

Despite their apparent variations, different attacks follow similar design principles, which entail invariant properties shared by adversarial inputs, independent of concrete attacks. In specific, we exploit the
{\em minimality principle} underlying most attack models as the foundations for building attack-agnostic defenses.

Intuitively, to maximize the attack's evasiveness, the adversary often seeks the minimum possible distortion to convert genuine inputs to adversarial ones. Formally,
\begin{definition}[Minimality Principle]
Given the target \dnn $f$, genuine input $\mbx$, and adversarial output $\mbyp$, the attack seeks to solve an optimization problem as:
\begin{equation*}
\min_{\mbr} ||\mbr|| \quad \textrm{s.t.}\quad f(\mbx + \mbr) = \mbyp
\end{equation*}
\end{definition}
For example, \cite{Goodfellow:2014:arxiv,Huang:2015:arxiv} instantiate this problem with $||\cdot||$ defined as $l_\infty$-norm, while \cite{Papernot:2016:eurosp,Carlini:2016:arXiv} consider $l_1$-norm.

%

\begin{figure}
\centering
\epsfig{file=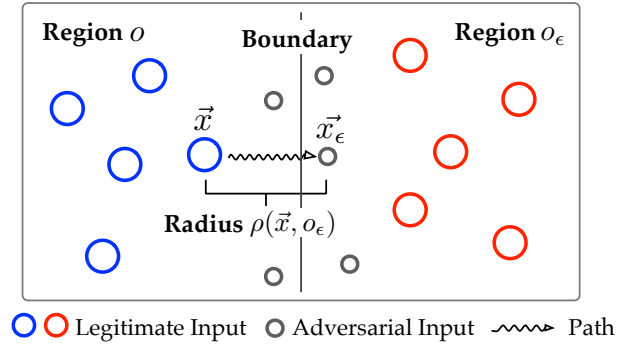, width=85mm}
\caption{Concepts of boundary, path, and radius. \label{fig:concept}}
\end{figure}

To understand the entailments of this principle, we first introduce several key concepts (see Figure~\ref{fig:concept}).

\begin{definition}[Boundary]
A \dnn $f$ partitions the input space (the topological space spanned by all the inputs) into non-overlapping regions. The inputs in each region are classified by $f$ into the same class. Adjacent regions are separated by their \underline{boundary}.
\end{definition}



\begin{definition}[Path]
For given inputs $\mbx$, $\mbxp$ with $\mbxp = \mbx + \mbr$, the \pv $\mbr$ encodes a \underline{path} from $\mbx$ to $\mbxp$, of which
the length is defined as the magnitude of $\mbr$, $||\mbr||$.
\end{definition}


\begin{definition}[Radius] The path length of an input $\mbx$ to its nearest neighbor $\mbxp$ in another class $\mbyp$
     is referred to as $\mbx$'s \underline{radius} to class $\mbyp$, denoted by $\rho (\mbx, \mbyp)$.
\end{definition}



%



We now translate the minimality principle in the language of boundary, path, and radius: given a genuine input $\mbx$, among all the possible (adversarial) inputs in the target class $\mbyp$, the adversary seeks $\mbxp$ with the shortest path length from $\mbx$. Therefore, the minimality principle entails the following important properties:
\begin{itemize}
\item Property 1: the path length of $\mbx$ to $\mbxp$ approximates $\mbx$'s radius to $\mbyp$, $\rho(\mbx, \mbyp)$.
\item Property 2: $\mbxp$ tends to distribute extremely close to the boundary of $\mby$ and $\mbyp$.
\end{itemize}

Next we empirically verify these properties, while their analytical treatment is deferred to~\myref{sec:analysis}.

Specifically, given a genuine input $\mbx$ (in class $\mby$) and an adversarial one $\mbxp$ (in class $\mbyp$) generated by an attack ${\mathcal A}$, in the given dataset, we find $\mbx$'s closest genuine counterpart $\mbxy$ in class $\mbyp$, i.e., $||\mbx - \mbxy||$ is minimized. We then compute the ratio of $\mbx$'s distance to $\mbxp$ and $\mbxp$:
$||\mbx - \mbxp||/||\mbx - \mbxy||$.

\begin{figure}
\epsfig{file=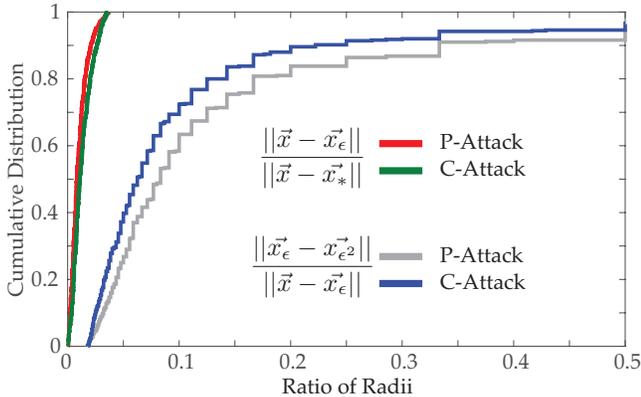, width=85mm}
\caption{Cumulative distribution of the ratio of input $\vec{x}$'s shortest distance to adversarial and genuine inputs (on the \svhn dataset). \label{fig:sparse}}
\end{figure}

Figure~\ref{fig:sparse}\; shows the cumulative distribution of such ratios with respect to \ttp and \ttca on the \cifar dataset (similar results observed on other datasets and attacks). Observe that most ratios lie in the interval of $[0, 0.01]$, regardless of attacks, suggesting
 that $\mbx$ resides much closer to its nearest adversarial neighbor in $\mbyp$ than to its genuine counterpart. Thus $||\mbx - \mbxp||$ approximates $\mbx$'s radius to $\mbyp$.

Further, by applying ${\mathcal A}$ to the adversarial input $\mbxp$, we generate another adversarial one\footnote{Strictly speaking, $\mbxpp$ is adversarial, given that it is created by perturbing another adversarial input $\mbxp$. Here we broadly refer to all the artificially generated inputs as adversarial inputs.} $\mbxpp$ in class $\mby$; similarly, $||\mbxp - \mbxpp||$ approximates $\mbxp$'s radius to $\mby$, $\rho(\mbxp, \mby)$. We then compute the quantity of $||\mbxp - \mbxpp||/||\mbx - \mbxp||$, i.e., a proxy for $\rho(\mbxp, \mby)/\rho(\mbx, \mbyp)$.

Figure~\ref{fig:sparse}\; shows the cumulative distribution of such ratios. Across both attacks, over 80\% of the ratios concentrate in the interval of $[0, 0.2]$, indicating $\mbxp$ distributes closer to the boundary than its genuine counterpart $\mbx$.

\begin{figure*}
\centering
\epsfig{file=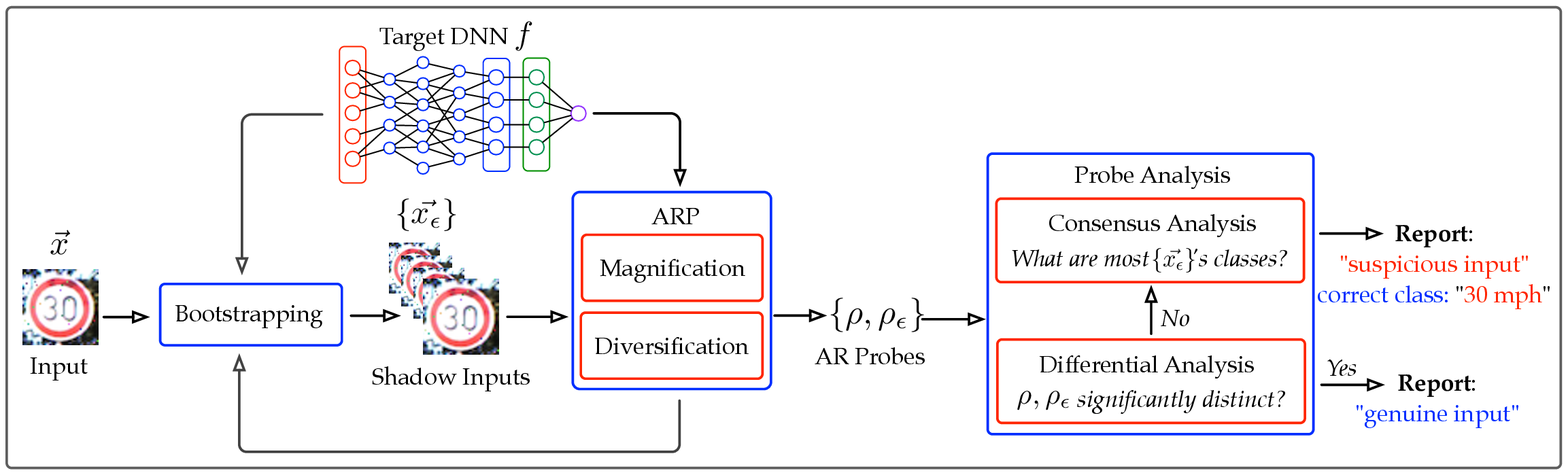, width=180mm}
\caption{Illustration of \system architecture. It discriminates adversarial inputs through the lens of adversarial radius analysis; for suspicious inputs, it further attempts to uncover their correct classification outputs. \label{fig:framework}}
\end{figure*}

\subsection{Building EagleEye}

These properties provide the premise for building effective differentiators to identify adversarial inputs: for a given input $\mbx$ (classified by $f$ as $\mby$), we measure its radii to all other classes, among which we find the minimum one: $\min_{\mbyp \neq \mby} \rho(\mbx, \mbyp)$, referred to as its {\em adversarial radius} (\ar).\footnote{Similar definitions have also been discussed in~\cite{Fawzi:2015:arxiv,Bastani:2016:arXiv,Feng:2016:arXiv}.} With Property 1 and 2, we can differentiate genuine and adversarial inputs via examining their \ars.

However, to realize this idea, we face two major challenges. First, the radius metrics are attack-specific, e.g., it is measured differently by \ttg and \ttpa. Directly measuring radii is at most effective for specific attacks. Second, even for known attacks, finding an optimal threshold is difficult; even if it exists, it tends to vary with concrete datasets and \dnn models.

To tackle the first challenge, we propose {\em adversarial radius probing} (\arp), an attack-neutral method to indirectly approximate \ar. In specific,
it employs semi-random perturbations and measures an input $\mbx$'s \ar as the minimum distortion amplitude (referred to as its \ar probe or probe) required to change its classification outputs.

To address the second challenge, we apply a bootstrapping method to remove the need for error-prone parameter tuning. In specific, for the given input $\mbx$, we generate a set of {\em shadow inputs} $\{\mbxp\}$ via semi-random perturbations. By comparing the probes of $\mbx$ and $\{\mbxp\}$ (differential analysis), we estimate the likelihood that $\mbx$ has been maliciously tampered. If suspicious, we further infer $\mbx$'s correct classification output by analyzing the consensus of $\{\mbxp\}$ (consensus analysis).

The framework of \system is illustrates in Figure~\ref{fig:framework}. Below we elaborate its key components,
\arp in \myref{sec:arp}, bootstrapping in \myref{sec:boostrap}, and probe analysis in \myref{sec:arpa}.

\subsubsection{Adversarial Radius Probing}
\label{sec:arp}

In this stage, by performing random perturbations on the given input $\mbx$, \system estimates the minimum distortion amplitude (i.e., probe) necessary to change its classification by $f$. Intuitively, $\mbx$'s probe, denoted by $\rho(\mbx)$, reflects its \ar in an attack-neutral manner.

Yet, it is often infeasible to estimate $\rho(\mbx)$ by running random perturbations on all of $\mbx$'s components given its high dimensionality. We use a semi-random perturbation method: (i) {\em magnification} - \system first dynamically identifies a set of {\em saliency regions} in $\mbx$ that maximally impact its classification; (ii) {\em diversification} - it performs random perturbations over such regions to estimate
$\rho(\mbx)$. We detail these two operations below.

%
%
%
%

\paragraph*{Magnification.\;}
\label{sec:mag}

The magnification operation is loosely based on attention mechanisms~\cite{gregor:2015:arxiv,Almahairi:2016:icml}, inspired by that human's vision automatically focuses on certain regions of an image with ``high resolution'' while perceiving surrounding regions in ``low resolution''. We use a simple attention mechanism with computational advantages.



For the given input $\mbx$, we define a set of $d \times d$ spatial regions ($d=4$ in our implementation). We generate all possible regions of $\mbx$ by applying an identity kernel of size $d\times d$ over $\mbx$, similar to a convolution operation.

To select the top $n$ saliency regions that maximally impact $\mbx$'s classification, we apply a greedy approach.
We sort the components of $\mbx$ according to their saliency (e.g., its gradient or Jacobian value) in descending order. Let $\rm{rank}(x)$ be the ranking of the component $x$. The saliency of a region $\mbg$ is the aggregated saliency contributed by all the components contained in $\mbg$:
\begin{displaymath}
{\rm saliency}(\mbg) = \sum_{x \in \mbg} c^{-\rm{rank}(x)}
\end{displaymath}
where $c$ $(c \geq 1)$ is a constant.

This definition allows us to control the characteristics of selected regions. With large $c$, we focus on regions that cover the most influential components; while with small $c$ (i.e., close to 1), we find regions that contain less influential components. The rationale behind balancing these two factors is as follows: the saliency landscape shifts as $\mbx$ is perturbed as $\mbxp$; yet, due to the inherent continuity of \dnns, the variation tends to be local. By properly setting $c$, we are able to accommodate such shift and still capture the most influential component in $\mbxp$.

We iteratively select the top $n$ regions. Let $\mathcal{R}_i$ be the selected regions after the $i^{\rm th}$ iteration. We then update the saliency of each remaining region by removing the contributions by components contained in regions in $\mathcal{R}_i$. Formally,
\begin{displaymath}
    {\rm saliency}(\mbg) = \sum_{x \in \mbg \bigcap \not\exists \mbg' \in \mathcal{R}_i, x \in \mbg'} c^{-\rm{rank}(x)}
\end{displaymath}

We then pick the region with the largest saliency among the remaining ones. We will discuss the optimal setting of $k$ and $c$ in \myref{sec:evaluation}.


\paragraph*{Diversification.\;}
\label{sec:div}



At this step, given the saliency regions ${\mathcal R}$ of $\mbx$, we perform random perturbations on ${\mathcal R}$ to estimate $\mbx$'s probe $\rho(\mbx)$.

With a little abuse of notations, let $\mbg$ be the set of components contained in the regions of $\mathcal{R}$. At each run, following a predefined distribution $p$ (with parameter $\theta$), we randomly select a subset of components in $\mbg$ to construct the perturbation vector $\mbr$, denoted by $\mbr \leadsfrom_\theta \mbg$. A successful perturbation $\mbr$ results in $f(\mbx + \mbr) \neq \mby$, where $\mby$ is $\mbx$'s current classification output by $f$.

In our current implementation, we instantiate $\theta$ as a uniform distribution and assume the flipping perturbation which sets an input component to either fully-on (`1') or fully-off (`-1'). The distortion amplitude is therefore measurable by the sampling rate $\theta$ of $p$. While other instantiations are certainly possible (e.g., zipf distribution), we find that this instantiation is both (i) effective in discriminating adversarial inputs (\myref{sec:evaluation}) and (ii) simple to control by the system operator. Moreover, the large entropy of uniform distributions enhances the hardness for the adversary to evade the detection (\myref{sec:analysis} and \myref{sec:evaluation}).

In the following, we consider the minimum sampling rate $\theta^*$ required to cause successful perturbations as $\mbx$'s probe, which indirectly reflects $\mbx$'s \ar.

\subsubsection{Bootstrap Operation}
\label{sec:boostrap}

By randomly perturbing the given input $\mbx$, the bootstrap operation produces a set of adversarial inputs $\{\mbxp\}$, which we refer to as $\mbx$'s {\em shadow inputs}. Intuitively, such shadow inputs represent $\mbx$'s near (if not the nearest) adversarial counterparts in other classes.


Specifically, to generate $\{\mbxp\}$, we adopt the same semi-random perturbation strategy as in \myref{sec:arp}, except for that the sampling rate is now fixed to be $\mbx$'s probe $\theta^*$. This practice ensures that the generated shadow inputs are as close to $\mbx$ as possible. Further, as will be revealed in \myref{sec:arpa}, in consensus analysis, this practice also helps uncover the correct classification of $\mbx$ if it is adversarial. To provide stable estimation, we require the number of shadow inputs to be larger than a threshold $k$ ($k \geq 4$ seems sufficient in practice \myref{sec:evaluation}).

We then estimate the probes of all the shadow inputs, $\{\rho(\mbxp)\}$, which, together with $\rho(\mbx)$, are fed as inputs to the phase of probe analysis. For simplicity of presentation, in the following, let $\rho$ and $\rho_\epsilon$ denote the probes of $\mbx$ and $\mbxp \in \{\mbxp\}$ respectively.

\subsubsection{Probe Analysis}
\label{sec:arpa}

In this phase, by analyzing the probes of the given input $\mbx$ and its shadow inputs $\{\mbxp\}$, \system determines the likelihood that $\mbx$ has been maliciously tampered (differential analysis); if so, it further attempts to recover its correct classification result (consensus analysis).

\paragraph*{Differential Analysis.\;} Recall that $\{\mbxp\}$ represent $\mbx$'s near (if not the nearest) adversarial neighbors in other classes. Thus, if $\mbx$ itself is adversarial, $\mbx$ and $\mbxp$ can be considered as adversarial versions of each other, thereby featuring similar \ars; otherwise, if $\mbx$ is a genuine input, the \ars of $\mbx$ and $\{\mbxp\}$ tend to show significant difference (see Figure~\ref{fig:sparse}).

In differential analysis, we leverage this insight and examine the probes of $\mbx$ and each shadow input $\mbxp$. Intuitively, a larger probe ratio of $\rho/\rho_\epsilon$ indicates that $\mbx$ is more likely to be genuine. Concretely, with a given shadow input $\mbxp$, we estimates the likelihood that $\mbxp$ is genuine as:
\begin{equation*}
{\rm genuine}_{\mbxp}(\mbx) = \frac{1}{1 + \exp(1-\rho/\rho_\epsilon)}
\end{equation*}
Here the sigmoid function converts $\rho/\rho_\epsilon$ to the interval of $(0, 1)$, which we may roughly interpret as the ``probability'' that $\mbx$ is genuine. In specific, this probability is 0.5 if $\rho/\rho_\epsilon = 1$. The overall likelihood that $\mbx$ is genuine is computed by aggregating the results over all the shadow inputs:
${\rm genuine}(\mbx) = \sum_{\mbxp}{\rm genuine}_{\mbxp}(\mbx)/|\{\mbxp\}|$.

In our empirical evaluation in \myref{sec:evaluation}, we find a threshold 0.625 works well across all the known attacks.

\paragraph*{Consensus Analysis.\;}
As shown in Figure~\ref{fig:framework}, if $\mbx$ passes the differential analysis, it is reported as ``genuine''; otherwise, it is considered as a ``suspicious'' case and moved to the phase of consensus analysis, in which we attempt to infer its correct classification output.

To be specific, recall that an adversarial input $\mbx$ (in class $\mbyp$) created from a genuine input (in class $\mby$) resides near to the boundary of $\mby$ and $\mbyp$. Thus, among its close adversarial neighbors in other classes, a majority of them should belong to class $\mby$. By leveraging this observation, we simply pick the most common class associated with the shadow inputs $\{\mbxp\}$ as $\mbx$'s most likely correct classification output.

%% file: analysis.tex
\section{Analysis of EagleEye}
\label{sec:analysis}

In the preceding sections, we present \system that applies adversarial radius analysis (\mar) to distinguish genuine and tampered inputs. Next, we analytically explore the effectiveness of \mar. Note that we do not intend to provide a definitive argument about using \system (or \mar) to mitigate adversarial input attacks, but rather we view it as an initial step towards building universal, attack-agnostic defenses against adversarial inputs for \dl and machine learning systems in general. In specific, our analysis attempts to draw the connection between \mar, \dnn generalizability, and learning theory.

Furthermore, we discuss the adversary's possible countermeasures to evade \system's detection. Recall that \system is built on the premise that the attacks follow the minimality principle: the adversary attempts to minimize the distortion amplitude to maximize the attack's evasiveness. We explore attack variants that (partially) abandon this principle. This amounts to investigating the adversary's design spectrum: the tradeoff between the evasiveness with respect to \system and the distortion amplitude (i.e., the evasiveness with respect to other detection mechanisms), which provides insights for the best practice of \system.


\subsection{Effectiveness of Radius Analysis}
\label{sec:mara}

The fundamental premise of \mar is that genuine inputs are inclined to have larger adversarial radii than their adversarial counterparts. Below we provide the theoretical underpinnings of this property. For simplicity of exposition, we exemplify $l_1$-norm as the measure of distortion amplitude, while our discussion generalizes to other metrics as well.

From the view of an adversarial input $\mbxp$, recall that attack ${\mathcal A}$ produces $\mbxp$ by carefully perturbing a ßgenuine input $\mbx$. Regardless of its concrete implementation, ${\mathcal A}$ is essentially designed to solve the optimization problem:
\begin{equation*}
    \min_{\mbxp} ||\mbxp - \mbx|| \quad \mathrm{s.t.} \quad f(\mbxp) = \mbyp
\end{equation*}
where $||\mbxp - \mbx||$ reflects the perturbation amplitude. Assume that ${\mathcal A}$ operates on $\mbx$ with a sequence of $t$ perturbations and $\mbr^{(i)}$ represent the perturbation vector at the end of the $i^{\mathrm{th}}$ iteration ($i = 1, 2, \ldots, t$), with $\mbxp =  \mbx + \mbr^{(t)}$. The minimality principle implies that ${\mathcal A}$ achieves the desired misclassification only after the $t^{\mathrm{th}}$ iteration, i.e.,
\begin{equation*}
\begin{array}{l}
f(\mbx + \mbr^{(t-1)}) = \mby\\
f(\mbx + \mbr^{(t)}) = \mbyp
\end{array}
\end{equation*}

Therefore, $(\mbx + \mbr^{(t-1)})$ and $(\mbx + \mbr^{(t)})$ represent two inputs lying between the class boundary of $\mby$ and $\mbyp$. It is also noted that $\mbr^{(t-1)}$ and $\mbr^{(t)}$ differ only by a few components, depending on the concrete implementation of ${\mathcal A}$. For example, the attacks in~\cite{Papernot:2016:eurosp,Carlini:2016:arXiv} perturb two input components at each round, and $\mbr^{(t-1)}$ and $\mbr^{(t-1)}$ differ by two components. We thus conclude that $\mbxp$ resides extremely close to the class boundary of $\mby$ and $\mbyp$.


Next, we move on to explain the effectiveness of \mar from the perspective of a genuine input $\mbx$. Our intuition is that
if $\mbx$ is classified as class $\mby$ by \dnn $f$ with high confidence, its radii to the class boundaries induced by $f$ must be reasonably large. To make this intuition more precise, we resort to statistical learning theory on the connection of classification confidence and \mar.


\begin{definition}[Confidence]
The classification \underline{confidence} of an input-output pair $(\mbx, \mby)$ (i.e., $f$ classifies $\mbx$ as $\mby$) is measured by the difference of the largest and second largest probabilities in $f(\mbx)$ (e.g., the softmax output). Formally,
$\phi(\mbx)  =  \min_{\mbyp \neq \mby} \sqrt{2}(\delta_{\mby} - \delta_{\mbyp}) \cdot f(\mbx)
$, where $\delta_i$ is the Kronecker delta vector with the $i^{\mathrm{th}}$ element being 1 and 0 otherwise and $\cdot$ denotes inner product.
\end{definition}

We now introduce an interesting result that connects the concept of classification confidence with \mar (adapted from~\cite{Sokolic:2016:arXiv} , pp. 14):

\begin{theorem*}
Given a \dnn $f$ and a genuine input $\mbx$, with $W^{(l)}$ being the weight matrix of the $l^{\mathrm{th}}$ layer of $f$, we have the following bound for the \ar of $\mbx$:
\begin{equation*}
\rho(\mbx) \geq  \frac{  \phi(\mbx) }{  \prod_{W^{(l)}} ||W^{(l)}||_F }
\end{equation*}
where $||\cdot||_F$ represents Frobenius norm.
\end{theorem*}

Intuitively, if a \dnn $f$ makes confident classification of $\mbx$, $\mbx$ tends to have a large \ar. Astute readers may point to the possibility of using confidence instead of \mar to adversarial inputs. However, note that high confidence is only one sufficient condition for large \ars; as observed in our empirical evaluation in~\myref{sec:evaluation} and previous work~\cite{Goodfellow:2014:arxiv}, there are adversarial inputs with high classification confidence but show small \ars.

Another implication of the above theorem is that increasing the classification confidence of \dnns is beneficial for discriminating adversarial inputs. Along this direction, defensive distillation~\cite{Papernot:2016:sp} is designed exactly for this purpose: by increasing the temperature $\tau$, it amplifies the probability difference in the outputs of \dnns.
In~\myref{sec:evaluation}, we empirically show the synergistic effects of integrating defensive distillation and \system.

\subsection{Adversary's Dilemma}

We now explore possible attack variants that attempt to evade \system's detection  or \mar in specific. Since \system is built on top of the minimality principle underlying varied attack models, one possible way to evade its detection is to (partially) abandon this principle in preforming adversarial perturbations.

Specifically, following \myref{sec:mara}, let $\mbr^{(i)}$ be the perturbation vector at the end of the $i^{\mathrm{th}}$ iteration and assume the adversary achieves the desired misclassification right after the $t^{\mathrm{th}}$ iteration:
$f(\mbx + \mbr^{(t)}) = \mbyp$. Now, instead of stopping here, the adversary keeps on perturbing $\mbx$, attempting to increase its \ar $\rho(\mbxp)$. Even in the ideal case, the adversary needs to perturb at least $\rho(\mbx)$ input components, resulting in an approximate perturbation amplitude of $2\cdot\rho(\mbx)$. Intuitively, if $\mbx$ is selected at random, then the quantity of $2\cdot\rho(\mbx)$ represents the distance of two genuine inputs, while in real datasets, the difference of distinct inputs is fairly discernible even to human eyes. Furthermore, due to the high nonlinearity of \dnns, the extra perturbation amplitude is often much larger this lower bound. The empirical validation of this hypothesis is given in~\myref{sec:sec2}, in which we also consider the adversary's another countermeasure of random perturbations.

This analysis above reveals a difficult dilemma for the adversary: she desires to preserve the \ar of an adversarial input to evade \system's detection; yet, to do so, she is forced to introduce extra perturbations sufficient to transform one genuine input to another, thereby significantly reducing the attack's evasiveness regarding other detection mechanisms (e.g., human vision).

%% file: evaluation.tex
\section{Evaluation}
\label{sec:evaluation}

In this section, we empirically evaluate the efficacy of \system. Specifically, our experiments are designed to answer the following key questions.
\begin{itemize}
\item Q: {\em Does \system effectively distinguish adversarial and genuine inputs in an attack-agnostic manner?}

A: (\myref{sec:sec1}) \system achieves very high detection accuracy across benchmark datasets and attack models. For instance, its average recall and precision are 99.5\% and 97.0\% on the \mnist dataset. 
In particular, this performance is achieved under the same parameter setting without any tuning to datasets or attack models.

\item Q: {\em Does \system cause new attack-defense arm races?}

A: (\myref{sec:sec2}) To evade \system's detection, the adversary has to abandon the minimality principle by significantly increasing the distortion amplitude, which weakens the attack's evasiveness with respect to other detection mechanisms and has fairly low success rate (less than 40\% across benchmark datasets).

\item Q: {\em Does \system complement other defense solutions?}

A: (\myref{sec:sec3}) \system exerts minimal interference to existing system components, and is compatible with any defense mechanisms. Further, defense-enhanced \dnns, with stronger generalization capabilities, provide even better foundations for \system to operate, leading to above 4\% increase in both precision and recall.

\end{itemize}

\subsection{Experimental Settings}

We use the same set of original \dnn models and benchmark datasets as in the empirical study in~\myref{sec:measure}. The details of \dnn models and datasets are referred to Appendix A and B. We also evaluate \system against the set of representative attacks in~\myref{sec:attack}. The default setting of parameters is as follows: \# patches $n = 8$, ranking coefficient $c = 1.25$, and \# shadow inputs $k = 4$.


\subsection{Discriminant Power of EagleEye}
\label{sec:sec1}

In this set of experiments, we show that \system is capable of performing accurate detection of adversarial inputs in an  attack-agnostic manner. We prepare the testing set as follows. We first randomly sample 5,000 inputs from each dataset, which form the pool of genuine inputs. We then apply all the attacks to each genuine input to generate its adversarial versions (with the adversarial target class randomly selected and the perturbation amplitude limited as 112 for \ttp and \ttca and 0.25 for \ttg and \ttha as in~\cite{Goodfellow:2014:arxiv,Papernot:2016:sp}); those successfully crafted instances form the pool of adversarial inputs. Due to the high success rates of adversarial attacks (see Table~\ref{tab:robustness}), the genuine and adversarial pools are quite balanced.

We apply \system to detect adversarial inputs and use the following metrics to measure its performance:
\begin{displaymath}
\textrm{\small Recall} = \frac{tp}{tp + fn}  \quad \textrm{\small Precision} =  \frac{tp}{tp + fp}
\end{displaymath}
where $tp$, $fp$, and $fn$ represent the number of true positive, false positive, and false negative cases (adversarial: +, genuine: -). Intuitively, recall and precision measure the sensitivity and specificity of \system.

\begin{table}{\small
\begin{tabular}{|c|c|c|c|c|c|}
  \hline
  \multirow{2}{*}{\bf Dataset} & \multirow{2}{*}{\bf Metric} & \multicolumn{4}{c|}{\bf Attack Model}\\
   \cline{3-6}
&  &  {\bf G-} & {\bf H-} & {\bf P-} & {\bf C-} \\
    \hline
    \hline
 \multirow{2}{*}{\mnist}  & Precision &  95.0\% & 96.5\% & 98.4\% & 98.0\% \\
 & Recall &   99.2\% & 100.0\% & 99.2\% & 99.6\% \\
\hline
\hline
\multirow{2}{*}{\cifar}  & Precision &   88.8\% & 90.8\% & 89.9\% & 91.2\% \\
& Recall &   98.4\% & 94.4\% & 96.4\% & 99.6\% \\
\hline
\hline
\multirow{2}{*}{\svhn}  & Precision &   88.3\% & 88.6\% & 87.3\ & 88.2\% \\
& Recall &   99.2\% & 96.4\% & 99.2\% & 98.8\% \\
\hline
\end{tabular}
\caption{\system detection accuracy with respect to different benchmark datasets and attack models. \label{tab:dacc}}}
\end{table}

%
%
%

Table~\ref{tab:dacc} summarizes \system's performance against varied attacks on the benchmark datasets. Observe that \system provides universal, attack-agnostic protection for \dnns: across all the cases, \system achieves high precision (above 87\%) and recall (above 94\%), indicating its strong discriminant power against adversarial inputs. Note that in all these cases, \system has slightly better recall than precision.
We hypothesize that those false positive cases are genuine inputs with low classification confidence (see \myref{sec:analysis}), while in the simplest \mnist dataset, most inputs have fairly high confidence scores, resulting in its lowest false positive rates.

We also examine \system's impact on genuine cases misclassified by \dnns. For those cases, \system detects them as genuine inputs with accuracy of 96.4\%, 94.4\%, and 95.6\% on the \mnist, \cifar, and \svhn dataset respectively, implying that \system's performance is relatively independent of the \dnn's accuracy.


%

We then evaluate \system's effectiveness in uncovering the correct classification of adversarial inputs. Among the adversarial cases detected by \system, we calculate the recovery rate as the proportion of inputs whose classes are correctly inferred. We find \system's recovery is effective against both linear and nonlinear attacks. For example, against \ttha, it achieves 85.6\% recovery on the \mnist dataset; against \ttpa, it achieves 66.0\% recovery on the \cifar dataset. Note that this performance is achieved under the same parameter setting without any adjustment towards datasets or attacks; thus, we believe \system's performance can be further improved by fine parameter tuning. The experiments on parameter tuning are referred to Appendix E.

To summarize, in an attack-agnostic manner, \system effectively discriminates maliciously tampered inputs and even uncovers their original classification outputs. It can be seamlessly deployed into any existing \dl-powered systems. The analysis results of \system can be combined with the \dnn classification results to form comprehensive reports to enable more informative decision-making for the system operators.

\begin{table}{\small
    \centering
\begin{tabular}{|c|c|c|c|c|}
  \hline
  \multirow{2}{*}{\bf Datasets} & \multicolumn{4}{c|}{\bf Attack Models}\\
   \cline{2-5}
&  {\bf G-Attack} & {\bf H-Attack} & {\bf P-Attack} & {\bf C-Attack} \\
    \hline
    \hline
\mnist  & 62.2\% &  94.6\% & 61.2\% & 61.6\% \\
\hline
\cifar  & 70.0\% &  91.2\% & 42.8\% & 79.4\%\\
\hline
\svhn  & 74.8\% & 91.6\% & 27.2\% & 47.6\% \\
\hline
\end{tabular}
\caption{Failure rates of adversarial inputs to reach desirable \ars with respect to benchmark datasets and attacks. \label{tab:success}}}
\end{table}

\subsection{Adversary's Countermeasures}
\label{sec:sec2}

\begin{figure}
\hspace{-15pt}
\epsfig{file=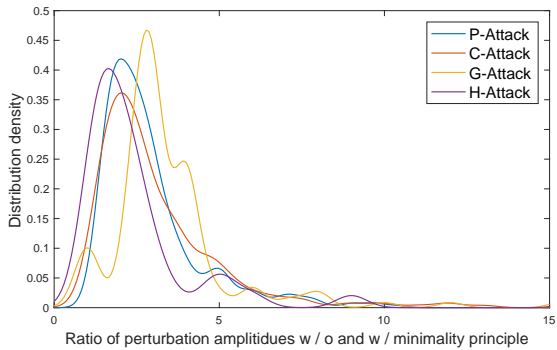, width=85mm}
\caption{Distribution of ratio of distortion amplitudes (without vs. with minimality principle) on \svhn. \label{fig:pratio}}
\end{figure}

Next we discuss adversary's possible countermeasures to evade \system's detection and their implications. Remember that the fundamental cornerstone of \system is the minimality principle underlying varied adversarial input attacks. Therefore, to evade \system's detection, one natural option for the adversary is to abandon this principle. Rather than applying the minimum possible distortion, she attempts to find a suboptimal perturbation vector leading to larger \ars.
%
%
%

Specifically, after the adversarial input $\mbxp$ achieves the misclassification $\mbyp$, the adversary continues the perturbation process, in an attempt to increase $\mbxp$'s \ar. Following the analysis in~\myref{sec:analysis}, here we empirically evaluate the implications of the adversary's countermeasure. Assume that for a given genuine input $\mbx$, the adversary desires to make $\mbxp$ with \ar comparable with that of $\mbx$.

We first investigate the cases that fail to achieve the desired \ar under the given perturbation amplitude, which is increased to 448 for \ttp and \ttca, and 1 for \ttg and \ttha. Table~\ref{tab:success} lists the failure rates of adversarial inputs on each dataset. For example, even by quadrupling the distortion amplitude, over 40\% of the inputs cannot achieve \ars comparable with their genuine counterparts on \cifar. This is explained by that due to the highly nonlinear, nonconvex nature of \dnns, the \ar of an adversarial input is a nonlinear function of the distortion amplitude as well. Thus, solely increasing the amplitude does not necessarily lead to the desired \ar.

Moreover, we examine those cases that indeed reach the desired \ars. Let $\mbxp$ and $\mbxp'$ represent the adversarial input generated with and without the minimality principle. We measure the ratio of their distortion amplitudes. Figure~\ref{fig:pratio} plots the distribution of such ratios on the \svhn dataset (experiments on other datasets in Appendix E). Note that regardless of the concrete attacks, for a majority of adversarial inputs, in order to make them evasive against \system's detection, the adversary has to amplify the distortion amplitude by more than 2.5 times than that guided by the minimality principle. Such large distortion would be detectable by potential anomaly detection systems or even human vision~\cite{Papernot:2016:eurosp}. This empirical evidence also validates our analysis in~\myref{sec:analysis}.

\begin{figure}
\hspace{-15pt}
\epsfig{file=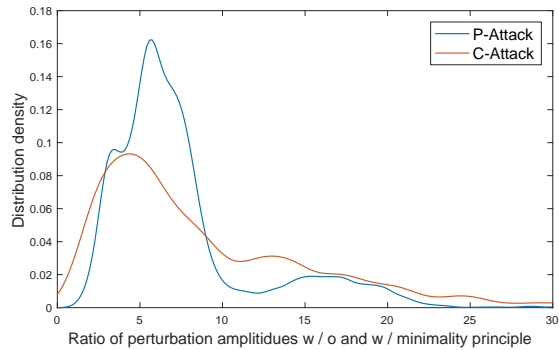, width=85mm}
\caption{Distribution of ratio of distortion amplitudes (random perturbation vs. minimality principle) on \svhn. \label{fig:pratio2}}
\end{figure}

Besides performing delicate perturbations to increase the \ars of adversarial inputs, the adversary may also try random perturbations in hope of finding adversarial inputs with satisfying \ars. We simulate this countermeasure by generating adversarial inputs with the minimum principle and then applying random perturbation to them. Figure~\ref{fig:pratio2} shows the ratio of distortion amplitudes after and before random perturbations (experiments on other datasets in Appendix E). Again, this countermeasure leads to significantly increased distortion amplitudes.

To summarize, \system creates a difficult dilemma for the adversary: she has to balance the attack's evasiveness against \system and other potential detection systems. In many cases, this balance is difficult to find, due to the highly nonlinear, nonconvex nature of \dnns.

%
%
%

\begin{table}{\small
    \centering
\begin{tabular}{|c|c|c|c|}
  \hline
  \multirow{2}{*}{\bf Cases} & \multicolumn{3}{c|}{\bf Dataset}\\
   \cline{2-4}
&  \mnist & \cifar & \svhn \\
    \hline
    \hline
defended by DD & 100.0\% &  98.4\% & 90.8\%\\
\hline
\hline
uncaptured by DD  & 100.0\% &  100.0\% &  100.0\% \\
\hline
\end{tabular}
\caption{Accuracy of \system against adversarial inputs defended/uncaptured by defensive distillation (DD).\label{tab:compat}}}
\end{table}

\subsection{EagleEye and Other Defenses}
\label{sec:sec3}

One advantage of \system is that it exerts minimal interference with existing system components. This feature makes it complementary with other defense mechanisms, while their integration often leads to synergistic effects. In this set of experiments, we validate this hypothesis by investigating the effects of applying \system on top of defensive distillation (\dd)~\cite{Papernot:2016:sp}.
As shown in Table~\ref{tab:robustness}, while effective against \ttg, \tth, and \ttpa, \dd is vulnerable to adaptively designed attacks (e.g., \ttca). For example, \ttca achieves near-perfect success rates on benchmark datasets.

First we consider adversarial inputs that are successfully defended by \dd. In particular, we apply \ttpa over original \dnns and collect all the adversarial inputs which pass original \dnns but are defended by \dd. For each adversarial input $\mbxp$ in this category, \dd recognizes its correct class but is unaware that $\mbxp$ is adversarial.
The first row of Table~\ref{tab:compat} lists \system's accuracy of detecting such cases as adversarial. It is clear that for cases successfully defended by \dd, \system can provide additional diagnosis information for the system operator.

Second we consider adversarial inputs that penetrate the protection of \dd. In particular, we apply \ttca over defensive distilled \dnns and collect all the successfully generated adversarial inputs. Table~\ref{tab:compat} lists \system's accuracy of detecting such cases as adversarial. \system achieves perfect detection rate in this category. It is clear for cases that penetrate the protection of \dd, \system provides another safe net.

%

We are not arguing to replace exiting defenses with \system. Rather, we believe it is beneficial to integrate complementary defense mechanisms, which significantly sharpens the edge of vulnerability to adaptive attacks.

%% file: discussion.tex
\section{Discussion}
\label{sec:discussion}

The preceding analytical and empirical analysis shows that \system, adversarial radius analysis (\ara) in specific, effectively discriminates adversarial inputs and even reveals their correct classification outputs.

One limitation of \system is that its effectiveness, to some extent, depends on the generalization capabilities of \dnns, although practically useful \dnns need to have sufficient generalizability (\dnn generalizability and robustness are two related but distinct properties~\cite{Feng:2016:arXiv}). We thus argue that the research on improving \dnn generalizability and that on defense mechanisms against adversarial inputs complement each other. It is our ongoing research to improve the effectiveness of \system against ambiguous inputs and weak \dnn.


We measure the distortion amplitude using $l_1$- or $l_\infty$-norm. There are other metrics for measuring input distance. For example, crafting adversarial malware samples to evade malware detection may require adopting other metrics ~\cite{Fogla:2006:ccs,Grosse:arxiv:2016}. We plan to investigate how to extend our solution to other metrics and perturbations. Yet, we believe the minimality principle still holds. For example, the malware author still wishes to preserve malware's functional behaviors.

In~\myref{sec:evaluation}, we empirically show the synergistic effects of combining defensive distillation and \system. It is expected because defense-enhanced \dnns, with stronger generalization capabilities than original models, provide better foundations for \ara to operate. Thus, we consider the integration of other defense mechanisms (e.g., data augmentation and robust optimization) and \system as a promising future direction to explore.

%

 Finally, it is worth emphasizing that \system does not create new attack vectors. It can be deployed compatibly with existing defense solutions. Its premise, the minimality principle, is an underlying principle followed by many attack models~\cite{Goodfellow:2014:arxiv,Huang:2015:arxiv,Papernot:2016:eurosp,Carlini:2016:arXiv}. Even if the adversary knows that \system is deployed, the only way to evade its detection is to amplify the adversarial distortion amplitude, which however reduces the attack's evasiveness with respect to other defense mechanisms. Therefore, \system indeed creates a difficult dilemma for the adversary.

%% file: literature.tex
\section{Additional Related Work}
\label{sec:literature}

Next we review three categories of related work: adversarial machine learning, deep learning-specific attacks and defenses, and robustness of deep neural networks.

Lying at the core of many security-critical domains, machine learning systems are increasingly becoming the targets of malicious attacks~\cite{Barreno:2006:asiaccs,Huang:2011:aisec,Barreno:2010:SML}. Two primary threat models are considered in literature: (i) poisoning attacks, in which the attackers pollute the training data to eventually compromise the learning systems~\cite{Biggio:2012:icml,Xiao:2015:SVM,Rubinstein:2009:imc}, and (ii) evasion attacks, in which the attackers modify the input data at test time to trigger the learning systems to misbehave~\cite{Dalvi:2004:kdd,Lowd:2005:kdd,Nelson:2012:QSE}. Yet, for ease of analysis, most of the work assumes simple learning models (e.g., linear classifier, support vector machine, logistic regression) deployed in adversarial settings.

Addressing the vulnerabilities of deep learning systems to adversarial inputs is more challenging for they are designed to model highly nonlinear, nonconvex functions~\cite{Nguyen:2015:cvpr,Sabour:2016:iclr}. One line of work focuses on developing new attacks against \dnns~\cite{Goodfellow:2014:arxiv,Huang:2015:arxiv,Tabacof:2015:arXiv,Papernot:2016:eurosp,Carlini:2016:arXiv}, most of which attempt to find the minimum possible modifications to the input data to trigger the systems to misclassify.
The detailed discussion of representative attack models is given in~\myref{sec:attack}. Another line of work attempts to improve \dnns resilience against such adversarial attacks~\cite{Goodfellow:2014:arxiv,Gu:2014:arxiv,Huang:2015:arxiv,Papernot:2016:sp}. However, these defense mechanisms often require significant modifications to either \dnn architectures or training processes, which may negatively impact the classification accuracy of \dnns. Moreover, as shown in~\myref{sec:measure}, the defense-enhanced models, once deployed, can often be fooled by adaptively engineered inputs or by new attack variants. To our best knowledge, this work represents an initial step to attack-agnostic defenses against adversarial attacks.

Finally, another active line of research explores the theoretical underpinnings of \dnn robustness. For example,
Fawzi {\em et al.}~\cite{Fawzi:2015:arxiv} explore the inherent trade-off between \dnn capacity and robustness;
Feng {\em et al.}~\cite{Feng:2016:arXiv} seek to explain why neural nets may generalize well despite poor robustness properties;
and Tanay and Griffin~\cite{Tanay:2016:arxiv} offer a theoretical explanation for the abundance of adversarial inputs in the input manifold space.

%% file: conclusion.tex
\section{Conclusion}
\label{sec:conclusion}

In this paper, we presented a new approach to defend deep learning (\dl) systems against adversarial input attacks. Our work was motivated by the observations that the fundamental challenges to tackle adversarial inputs stem from their adaptive and variable nature, while static defenses can often be circumvented by adaptively engineered inputs or by new attack variants. We developed a principled approach that, by leveraging the underlying design principles shared by varied attacks, discriminates adversarial inputs in a universal, attack-agnostic manner. We designed and implemented \system, a prototype defense engine which can be readily deployed into {\em any} \dl systems, requiring no modification to existing components. Through comprehensive adversarial tampering analysis, \system enables more informative decision-making for the operators of deep learning systems. Our empirical evaluations showed that \system, when applied to three benchmark datasets, detected nearly 96\% adversarial inputs generated by a range of attacks.

%% file: appendix.tex
\section*{Appendix}

\subsection*{A. Datasets}

The \mnist dataset~\cite{mnist} constitutes a set of 28$\times$28 greyscale images of handwritten digits (`0'-`9'), with 60K training and 10K testing samples.

The \cifar dataset~\cite{cifar} consists of 32$\times$32 color images from ten classes (e.g., `airplane', `automobile', `bird'), split into 50K training and 10K testing samples.

The \svhn dataset~\cite{svhn} comprises color images of house numbers collected by Google Street View. We consider the format of 32$\times$32 pixel images. The task is to classify the digit around the center of each image. There are 73K and 26K digits in the training and testing sets, respectively.

All the datasets are centered and normalized such that the value of each pixel lies in the interval of $[-1, 1]$.

\subsection*{B. Implementation of DNN Models}
\label{sec:app:dnn}

In the following we present the detailed architectures of \dnn models used in our empirical evaluation.

\paragraph*{Convolutional Neural Network.\;}
The \dnn model applied to classifying the \mnist dataset is a convolutional neural network (\cnn). In particular, we adopt an architecture similar to~\cite{Papernot:2016:sp}, which is summarized in Table~\ref{tab:mnist}. In addition, we apply dropout (rate = 0.5) at both fully connected layers.

\begin{table}[h]{\small
  \centering
\begin{tabular}{c|c}
{\bf Layer} & {\bf Definition} \\
\hline
ReLU Convolutional & \# filters: 32, kernel: 3$\times$3   \\
\hline
ReLU Convolutional & \# filters: 32, kernel: 3$\times$3  \\
\hline
Max Pooling & pool: 2$\times$2, stride: 1\\
\hline
ReLU Convolutional & \# filters: 64, kernel: 3$\times$3  \\
\hline
ReLU Convolutional & \# filters: 64, kernel: 3$\times$3  \\
\hline
Max Pooling & pool: 2$\times$2, stride: 1\\
\hline
ReLU Fully Connected & \# units: 256 \\
\hline
ReLU Fully Connected & \# units: 256\\
\hline
Softmax & \# units: 10\\
\hline
\end{tabular}
\caption{Convolutional network architecture (\mnist) \label{tab:mnist} }}
\end{table}

\paragraph*{Maxout Network.\;}
The maxout network (\mxn) model generalizes conventional \cnn models by employing maxout activation functions, which pool over multiple affine feature maps in addition to pooling over adjacent spatial locations as in convolution operations. Therefore, an maxout convolutional layer is defined as the composition of one convolutional layer, one regular pooling layer, and one maxout pooling layer.

To classify the \cifar dataset, we adopt an architecture similar to that in~\cite{Goodfellow:2013:icml}, which consists of three maxout convolutional layers and one maxout fully connected layer, as detailed in Table~\ref{tab:cifar10}. We apply dropout (rate = 0.5) at each maxout convolutional layer.

\begin{table}[h]{\small
  \centering
\begin{tabular}{c|c}
{\bf Layer} & {\bf Definition} \\
\hline
ReLU Convolutional & \# filters: 128 kernel: 5$\times$5, padding: 4 \\
\hline
Max Pooling & pool: 3$\times$3, stride: 2 \\
\hline
Maxout & \# units: 64 (2 pieces/unit)\\
\hline
ReLU Convolutional & \# filters: 256 kernel: 5$\times$5, padding: 3 \\
\hline
Max Pooling & pool: 3$\times$3, stride: 2\\
\hline
Maxout & \# units: 128 (2 pieces/unit)\\
\hline
ReLU Convolutional &  \# filters: 256 kernel: 5$\times$5, padding: 3 \\
\hline
Max Pooling & pool: 3$\times$3, stride: 2\\
\hline
Maxout & \# units: 128 (2 pieces/unit)\\
\hline
ReLU Fully Connected & \# units: 2,000 \\
\hline
Maxout & \# units: 400 (5 pieces/unit)\\
\hline
Softmax & \# units: 10\\
\hline

\end{tabular}
\caption{Maxout network architecture (\cifar) \label{tab:cifar10} }}
\end{table}

\paragraph*{Network-in-Network.\;}

To classify the \svhn dataset, we apply an network-in-network (\nin) model, which features another distinct architecture. In conventional \cnn, the convolution filter is essentially a generalized linear model (\glm) for the underlying data patch. The \nin architecture replaces  \glm with an ``micro neural network'' structure which is a nonlinear function approximator to enhance the abstraction capability of the local model.

Following~\cite{lin:2014:iclr}, we use multilayer perceptron (\mlp) as the instantiation of micro network. In the resulting mlpconv layer, the \mlp is shared among all local receptive fields, while the feature maps are obtained by sliding the \mlp over the input in a similar manner as \cnn.

Specifically, our \nin model comprises three mlpconv layers, followed by one average pooling layer and one softmax layer, as summarized in Table~\ref{tab:svhn}.
Dropout (rate = 0.5) is applied at each mlpconv layer.

\begin{table}[h]{\small
  \centering
\begin{tabular}{c|c}
{\bf Layer} & {\bf Definition} \\
\hline
ReLU Convolutional & \# filters: 96, kernel: 5$\times$5, padding: 2 \\
\hline
ReLU Convolutional &  \# filters: 96, kernel: 1$\times$1 \\
\hline
Max Pooling & pool: 3$\times$3, stride: 2\\
\hline
\hline
ReLU Convolutional & \# filters: 192, kernel: 5$\times$5, padding: 2\\
\hline
ReLU Convolutional & \# filters: 192, kernel: 1$\times$1 \\
\hline
Max Pooling & pool: 3$\times$3, stride: 2\\
\hline
\hline
ReLU Convolutional & \# filters: 192, kernel: 3$\times$3, padding: 1 \\
\hline
ReLU Convolutional & \# filters: 192, kernel: 1$\times$1 \\
\hline
ReLU Convolutional & \# filters: 10, kernel: 1$\times$1 \\
\hline
Average Pooling & pool: 8$\times$8 \\
\hline
Softmax & \# units: 10\\
\hline

\end{tabular}
\caption{Network-in-network architecture (\svhn) \label{tab:svhn} }}
\end{table}

%

\subsection*{C. Implementation of Defense Methods}

We implement one representative defense mechanism from each category of defense strategies in~\myref{sec:defense}.

\paragraph*{Data Augmentation.\;} Recall that with data augmentation, adversarial inputs are incorporated in training a more robust \dnn $f'$. Our implementation of this defense mechanism proceeds as follows.
\begin{myitemize}
\item We begin with an initialized \dnn $f$ and an attack of interest ${\mathcal A}$;
\item At each iteration, regarding the current $f$, we apply ${\mathcal A}$ to a minibatch $\{(\mbx, \mby)\}$ randomly sampled from the training set, and generate an augmented minibatch $\{(\mbx, \mbxp, \mby)\}$;
%
We update $f$ using the objective function:
\begin{equation*}
\min_{f} \sum_{(\mbx, \mbxp, \mby)}  (\ell(f(\mbx), \mby) +  \ell(f(\mbxp), \mby))
\end{equation*}
\end{myitemize}
We instantiate the attack ${\mathcal A}$ as each of \ttg, \tth, \ttp, \ttca and refer to the resulting models as \ttg, \tth, \ttp, and {\ttct} \dnn respectively.

\paragraph*{Robust Optimization.\;} Recall that with robust optimization, one improves \dnn stability by preparing it for the worst-case inputs, which is often formulated as an minimax optimization framework.

We implement the framework in~\cite{Shaham:2015:arxiv}, wherein the loss function is minimized over adversarial inputs generated at each parameter update. To be specific, it instantiates Eqn.\;(\ref{eq:rof}) with finding the perturbation vector $\mbr$ (with respect to an input instance $(\mbx, o)$) that maximizes its inner product with the gradient of objective function:
\begin{equation}
    \label{eq:rof2}
    \mbr = \arg\max_{\mbr} \langle \triangledown \ell (f(\mbx), o)), \mbr \rangle
\end{equation}
For $l_\infty$-norm, we limit $||\mbr||_\infty \leq 0.2$; for $l_1$-norm, we limit $||\mbr||_1 \leq 2*2$ (no more than 2 pixels).

\paragraph*{Model Transfer.\;} Recall that with model transfer, the knowledge in a teacher \dnn $f$ (trained on legitimate inputs) is extracted and transferred to a student \dnn $f'$, such that $f'$ generalizes better to adversarial inputs. Specifically, we implement the defensive distillation mechanism in~\cite{Papernot:2016:sp} as a representative method of model transfer. We set the temperature $\tau = 40$ in the experiments as suggested by~\cite{Papernot:2016:sp}.

\subsection*{D: Training of DNN Models}

The training process identifies the optimal setting for a \dnn's parameters $\mbw$. Due to complex structures of \dnn models and massive amount of training data,
we apply Stochastic Gradient Descent with Nesterov momentum~\cite{Sutskever:2013:icml} as the optimization algorithm. More specifically, let $\ell(\mbw)$ represent the objective function (e.g., the cross entropy of ground-truth class labels and \dnn models' outputs). At each epoch, the gradient of $\ell$ with respect to $\mbw$ is computed over a ``mini-batch'' (e.g., each of 128 samples) sampled from the training data via a back-propagation procedure. The update rule of $\mbw$ is given by:
\begin{equation*}
\left\{
\begin{array}{l}
v = \mu v -\lambda \cdot \triangledown \ell(\mbw + \mu v)\\
\mbw := \mbw + v
\end{array}
\right.
\end{equation*}
where $v$ represents the ``velocity'', $\mu$ denotes the ``momentum'', $\lambda$ is the learning rate, and $\triangledown$ is the gradient operator. The training process repeats until the objective function converges.

In implementation, the default learning rates for \mnist, \cifar, and \svhn datasets are respectively set as 0.1, 0.01, and 0.01; the default momentum $\mu$ is fixed as 0.9; the optimization algorithm is run for up to 240 epochs. In addition, we apply an adaptive learning rate scheme: at each epoch, let $\ell$ and $\ell^*$ respectively be the loss of current epoch and the best loss thus far; the learning rate is adjusted as: $\lambda = \lambda \cdot \exp (\frac{\ell^* - \ell}{s})$, where $s = 2.5$ if $\ell^* \geq \ell$ and $s = 0.75$ if  $\ell_* < \ell$. The \dnn models are trained using the training set of each dataset.
 All the algorithms are implemented on top of Theano\footnote{Theano: http://deeplearning.net/software/theano/}, a Python-based \dl library. All the experiments are performed using an array of 4 Nvidia GTX 1080 GPUs.

\subsection*{E: Additional Experiments}

Here we list the experiment results in addition to those in \myref{sec:evaluation}, including the impact of parameter tuning on \system's performance, and the tradeoff between the attack's evasiveness regarding \system and other defenses.

\begin{figure}
\epsfig{file=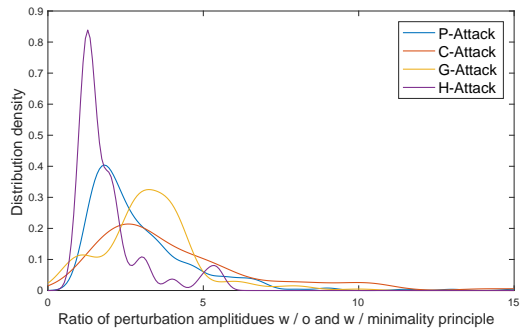, width=80mm}
\caption{Distribution of ratio of distortion amplitudes (without vs. with minimality principle) on \cifar. \label{fig:pratiocifar}}
\end{figure}

\begin{figure}
\epsfig{file=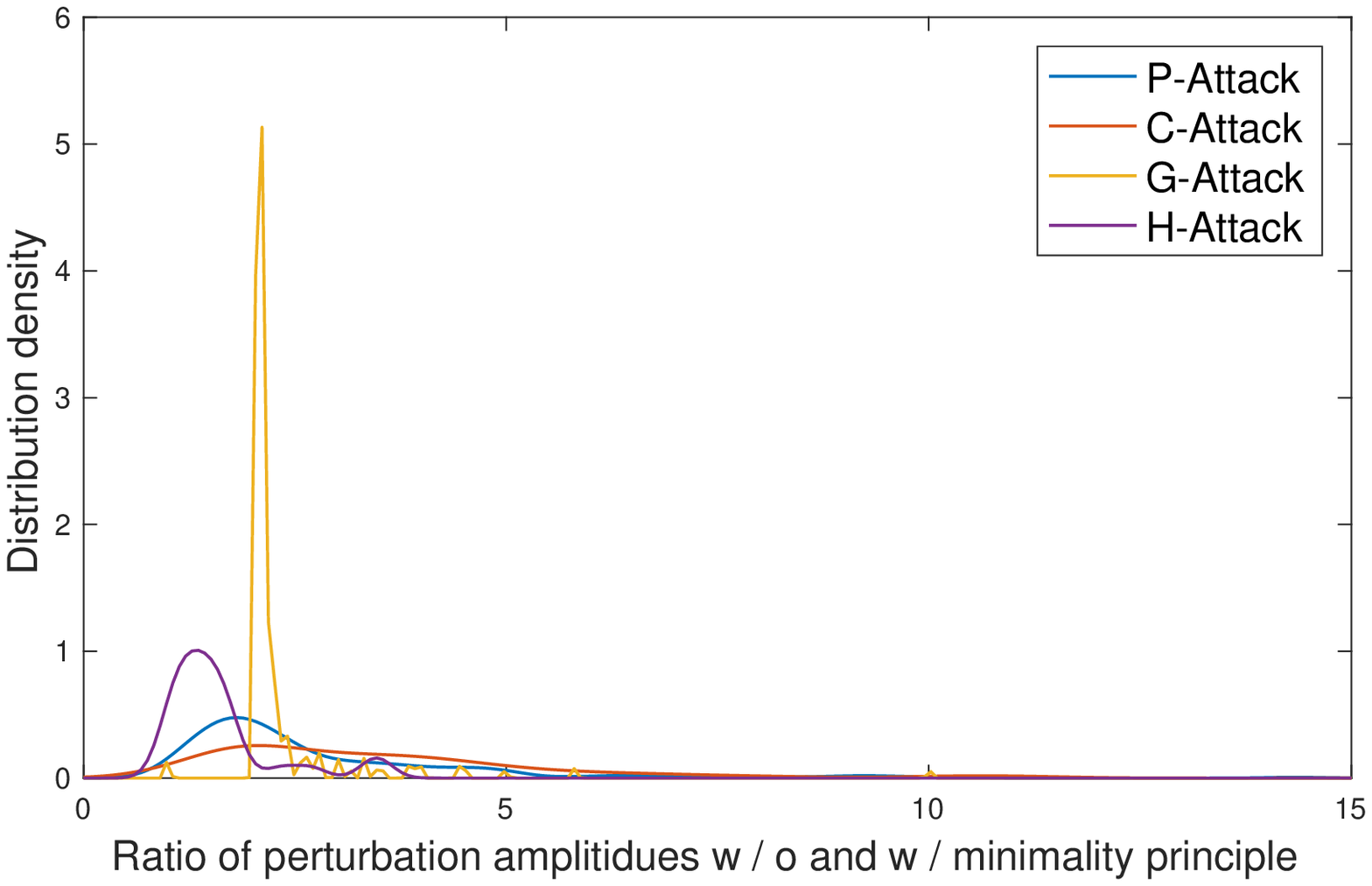, width=80mm}
\caption{Distribution of ratio of distortion amplitudes (without vs. with minimality principle) on \mnist. \label{fig:pratiomnist}}
\end{figure}

\begin{figure}
\epsfig{file=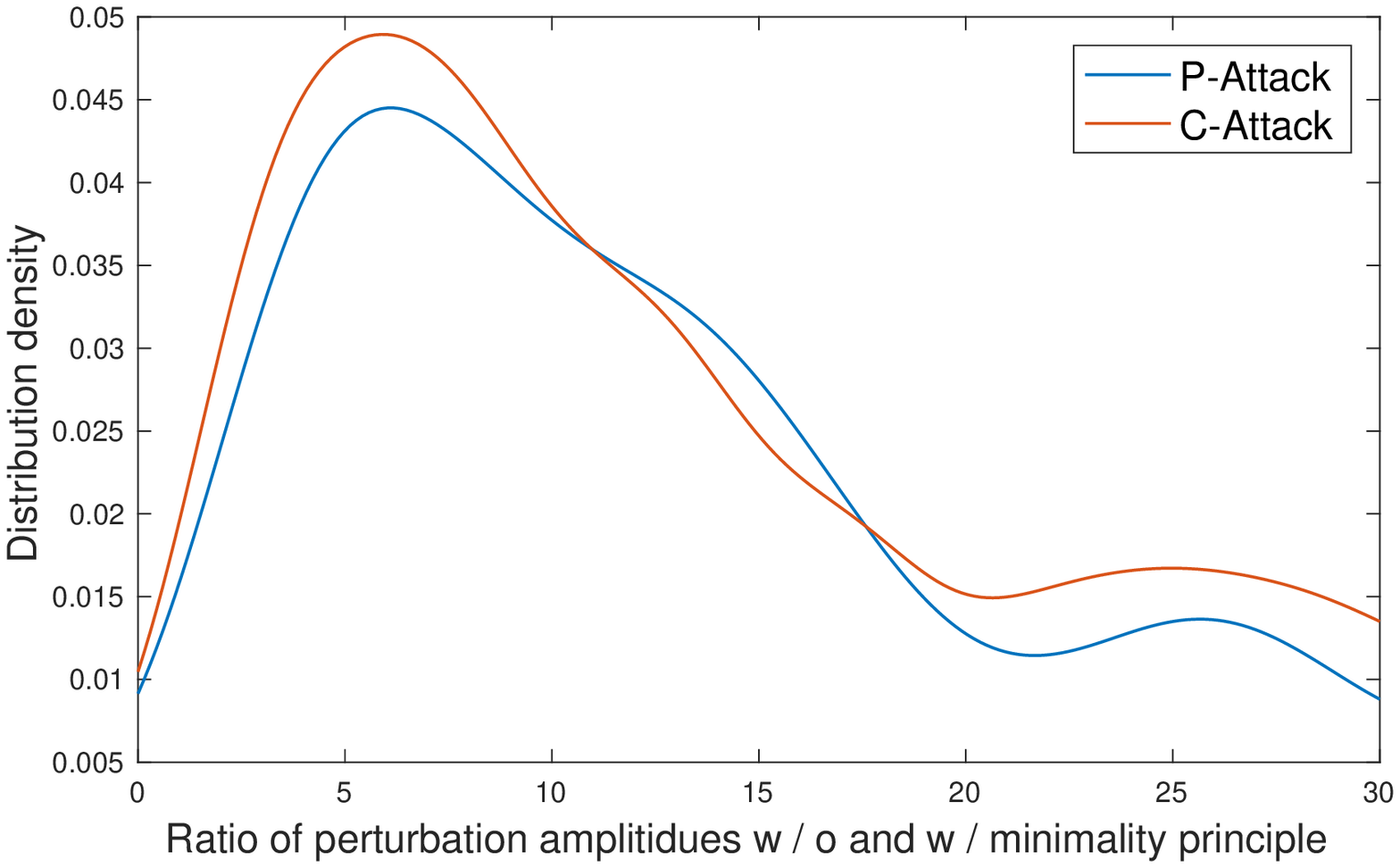, width=80mm}
\caption{Distribution of ratio of distortion amplitudes (random perturbation vs. minimality principle) on \cifar. \label{fig:pratiocifar2}}
\end{figure}

\begin{figure}
\epsfig{file=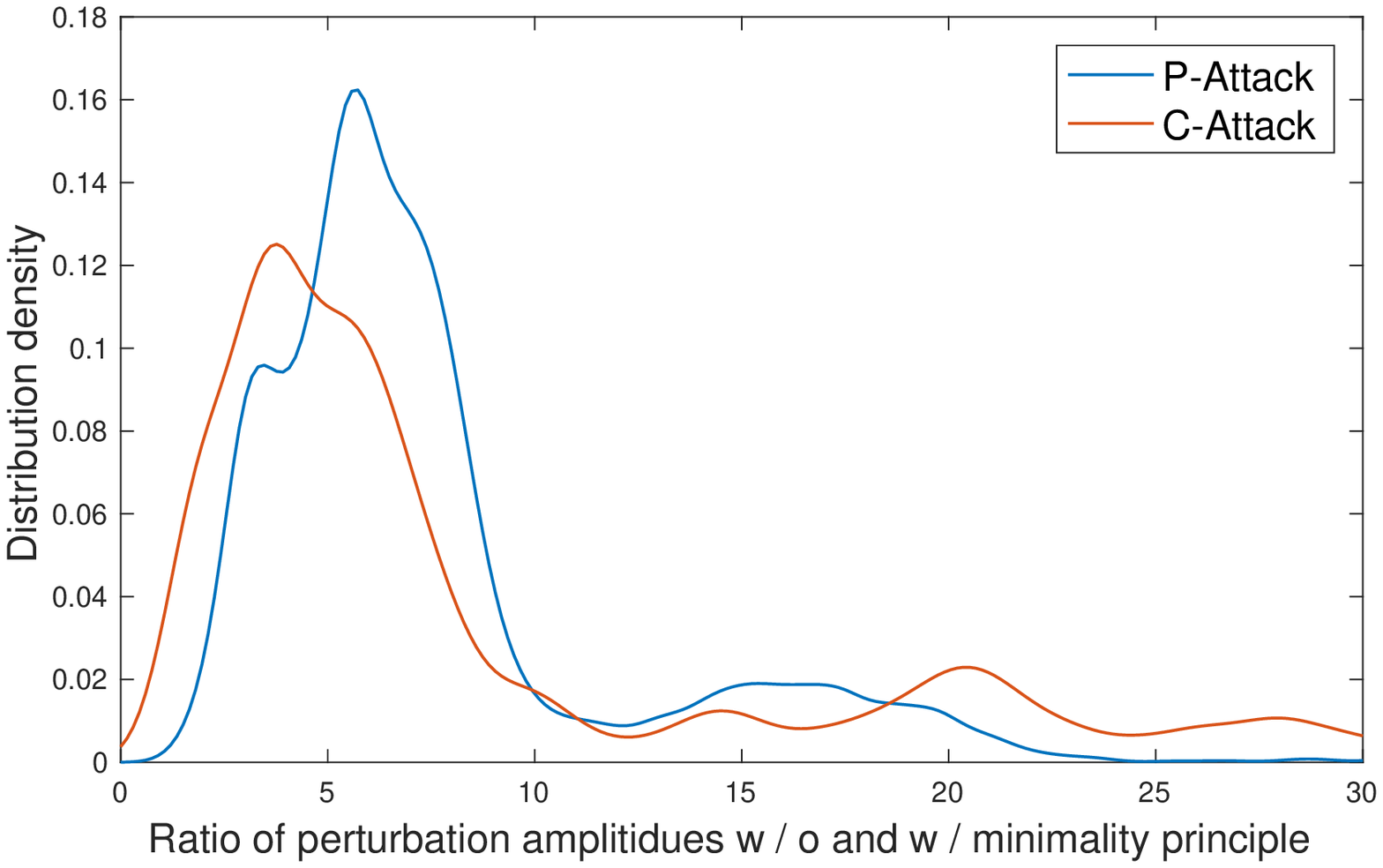, width=80mm}
\caption{Distribution of ratio of distortion amplitudes (random perturbation vs. minimality principle)  on \mnist. \label{fig:pratiomnist2}}
\end{figure}

\begin{figure}
\epsfig{file=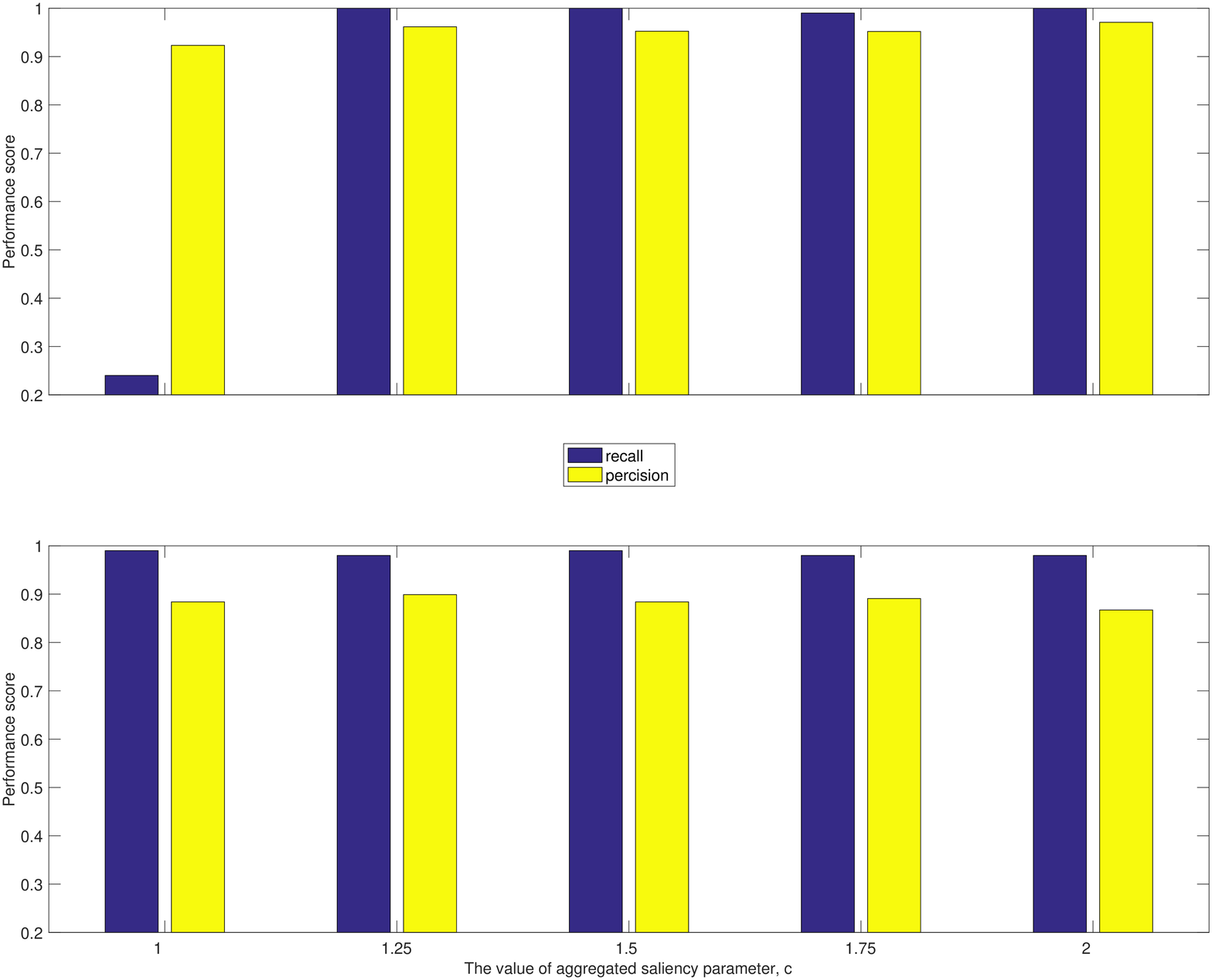, width=80mm}
\caption{Impact of number of patches $c$ on \system's performance. \label{fig:coefc}}
\end{figure}

\begin{figure}
\epsfig{file=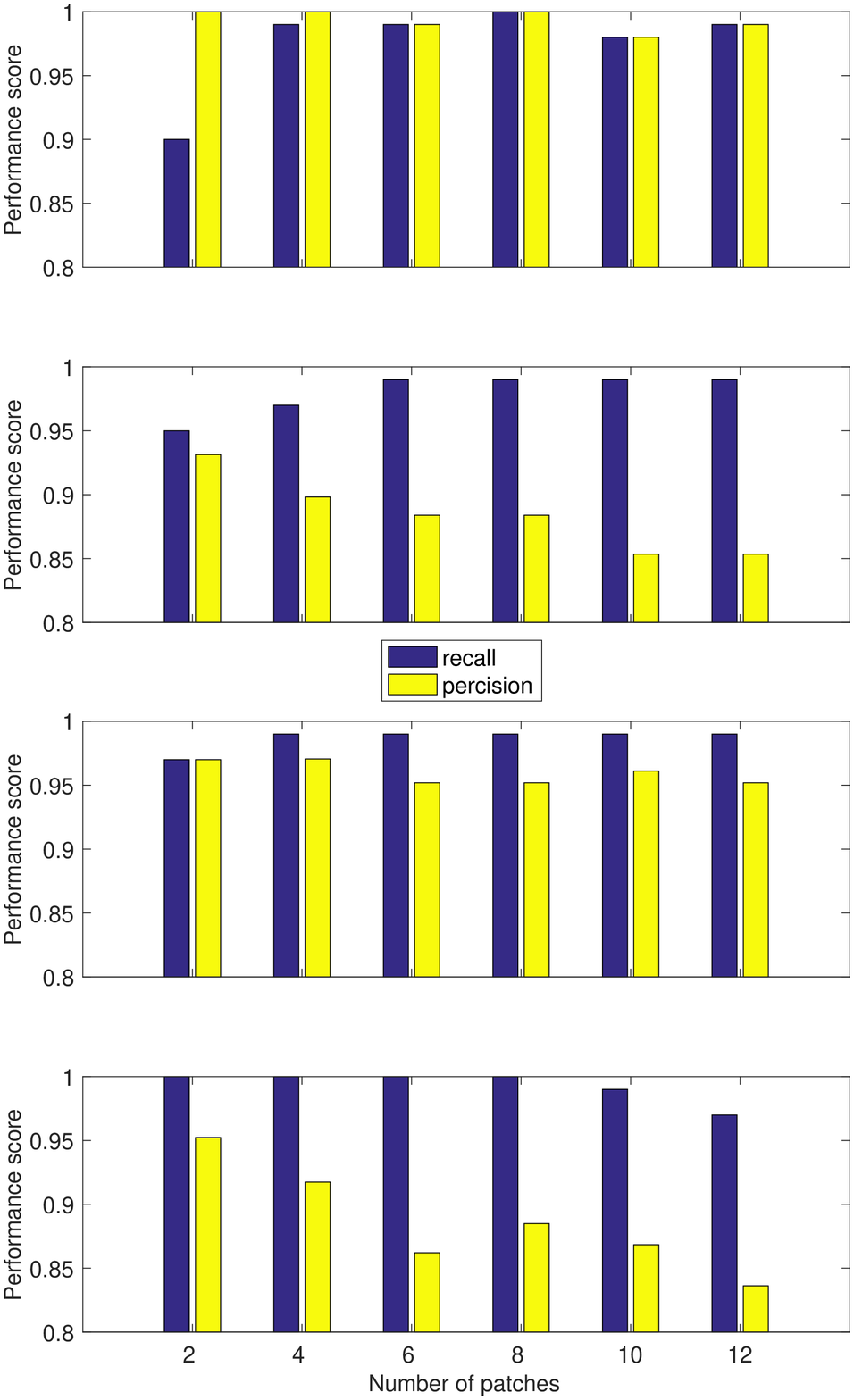, width=80mm}
\caption{Impact of number of patches $n$ on \system's performance. \label{fig:coefn}}
\end{figure}